\documentclass[12pt,a4paper]{article}
\usepackage{a4wide}

\usepackage[english]{babel}
\usepackage[utf8]{inputenc}

\usepackage{graphicx}
\usepackage{axodraw}
\usepackage[numbers,sort&compress]{natbib}
\usepackage{amsfonts,amsmath,amssymb}
\usepackage[usenames,dvipsnames]{xcolor}
\usepackage[ps2pdf,bookmarks,colorlinks,bookmarksnumbered,citecolor=OliveGreen]{hyperref}

\allowdisplaybreaks[4]

\newcommand{\phys}{\mathrm{phys}}
\newcommand{\tree}{\mathrm{tree}}
\newcommand{\tr}[1]{\langle #1 \rangle}

\renewcommand{\L}{\mathcal{L}}
\renewcommand{\O}{\mathcal{O}}

\newcommand{\M}{\mathcal{M}}

\newcommand{\unitmatrix}{\mbox{\boldmath$1$}}

\begin{document}

\begin{titlepage}
\begin{flushright}
\small
LU TP 13-10\\
revised April 2013\\
\end{flushright}

\vfill

\begin{center}
{\LARGE\bf Leading logarithms in $N$-flavour mesonic\\[3mm]Chiral Perturbation Theory}
\\[2cm]
{\bf Johan Bijnens$^a$, Karol Kampf$\hskip0.23ex^b$ and Stefan
Lanz$^{a,c}$}\\[4mm]
{$^a\,$Department of Astronomy and Theoretical Physics, Lund University,}\\
{S\"olvegatan 14A, S 223 62 Lund, Sweden}\\[3mm]
{$^b\,$Institute of Particle and Nuclear Physics, Faculty of Mathematics and
Physics,}\\
{Charles University, V Holesovickach 2, CZ-18000 Prague, Czech
Republic.}\\[3mm]
{$^c\,$Albert Einstein Center for Fundamental Physics, Institute for Theoretical Physics,}\\
{University of Bern, Sidlerstrasse 5, CH-3012 Bern, Switzerland.}
\end{center}

\vfill

\begin{abstract}
We extend earlier work on leading logarithms in the massive
nonlinear $O(n)$ sigma model to the case of $SU(N)\times SU(N)/SU(N)$
which coincides with mesonic chiral perturbation theory for $N$ flavours 
of light quarks. We discuss the leading logarithms for the mass and
decay constant to six loops and for the vacuum expectation value
$\langle \bar q q\rangle$ to seven loops. For dynamical quantities the
expressions grow extremely large much faster such that we only quote
the leading logarithms to five loops for the vector and scalar form factor
and for meson-meson scattering.
The last quantity we consider is the vector-vector to meson-meson
amplitude where we quote results up to four loops for a subset
of quantities, in particular for the pion polarizabilities.
As a side result we provide an elementary proof that the
factors of $N$ appearing at each loop order are odd or even depending on the
order and the remaining traces over external flavours.

\vspace{3ex}

\end{abstract}

\vfill

\end{titlepage}

\tableofcontents

\section{Introduction}

The calculation of higher loop corrections is an important problem in
all areas of particle physics. The leading logarithms in a renormalizable
field theory can be calculated to all orders by simply using the renormalization group.
In nonrenormalizable effective field theories like Chiral Perturbation Theory (ChPT)
\cite{Weinberg:1978kz,Gasser:1983yg,Gasser:1984gg}, the recursive argument
underlying the renormalization group does not work since one has a new
Lagrangian at each order. Weinberg~\cite{Weinberg:1978kz}
showed that using the requirement that all nonlocal divergences cancel,
one could obtain the leading logarithms (LL) at two-loop order with only
one-loop calculations. This method has then been applied to various
processes at the two-loop level~\cite{Bijnens:1998yu}. That it works to all
orders was later proven using beta-functions~\cite{Buchler:2003vw} and also
with a more diagrammatic method~\cite{Bijnens:2009zi}.

Using this method,
\cite{Kivel:2008mf,Kivel:2009az,Koschinski:2010mr,Polyakov:2010pt} found
recursion relations valid in the massless limit and applied them to a number
of processes. Away from the massless limit the tadpoles do not vanish and this
causes the number of needed one-loop diagrams at every order to increase
considerably. A systematic method to automatize the calculational process was found in
\cite{Bijnens:2009zi} and then applied to a number of processes in the
normal \cite{Bijnens:2009zi,Bijnens:2010xg} and abnormal or anomalous
intrinsic parity \cite{Bijnens:2012hf} sector of the massive
$O(n)$ nonlinear sigma model. In the present paper we extend the calculations
in the even sector to the symmetry breaking pattern of
$SU(N)\times SU(N)/SU(N)$. All results are for the case of equal masses.

We discuss the leading logarithm contribution to the mass,
decay constant, and vacuum expectation value to sixth or seventh order.
Numerical results are discussed for the two physical cases $N=2,3$.
For the vector and scalar form factors we give expressions for the full
results and for the radius and curvature. We present no numerical results,
but some discussion of numerics for the scalar form factor for $N=2$
can be found in \cite{Bijnens:2010xg}. For meson-meson scattering we
present analytic results for the amplitude and the scattering lengths
up to fifth order. We show numerical results only for the singlet
scattering length for $N=3$, which we compare with the full two-loop
calculation as well. For $\gamma\gamma\to\pi\pi$ we give analytic
results for the full amplitude for general $N$ and for the polarizabilities
for $N=2$. For the latter we also present numerical results.

We provide some references to the $N=2$ and $N=3$ cases at the two-loop level
where the general-$N$ case is not known to that level. An extensive discussion
of the corresponding literature can be found in the review \cite{Bijnens:2006zp}.

In Sect.~\ref{sigmamodel} we present the model and the different
parametrizations we use. Sect.~\ref{sec:LLeven} describes the changes needed
compared to the $O(n)$ work and provides the necessary definitions such that the formulas
in this paper are self-contained. We do however not discuss in detail the methods used.
The remaining sections present results for the mass (Sect.~\ref{secmass}),
decay constant (Sect.~\ref{secdecay}), vacuum expectation value
(Sect.~\ref{vev}), vector form factor (Sect.~\ref{secvff}),
scalar form factor (Sect.~\ref{secsff}), 
meson-meson scattering (Sect.~\ref{secpipi})
and  vector-vector to meson-meson scattering (Sect.~\ref{ggpipi}).
In addition we prove in Appendix~\ref{appA} that only certain powers of $N$ can show
up at each order.

\section{\texorpdfstring{$N$}{N}-flavour mesonic Chiral Perturbation Theory}
\label{sigmamodel}

The Lagrangian of the massive nonlinear $SU(N)\times SU(N)/SU(N)$ sigma model
or $N$-flavour mesonic ChPT at lowest order is given by
\begin{align}
\L = \frac{F^2}{4} \tr{D_\mu U D^\mu U^\dagger}
 + \frac{F^2}{4} \tr{\chi U^\dagger + U \chi^\dagger} \; ,
\label{lagrangian}
\end{align}
where $U$ is a special unitary $N\times N$ matrix, which contains
 $N^2-1$ degrees of freedom. $\langle A \rangle = \mathrm{tr}(A)$.
The interaction with external axial-vector and vector fields enters
through the covariant derivative
\begin{align}
D_\mu U = \partial_\mu U - \frac{i}{2} [v_\mu,U] - \frac{i}{2} \{a_\mu,U\} \; ,
\end{align}
while the explicitly chiral symmetry breaking terms as well as the scalar
and pseudoscalar external sources are contained in
\begin{align}
\chi = 2 B (s+ip) + M^2\,\unitmatrix \; .		\label{chiDef}
\end{align}
The chiral $SU(N) \times SU(N)$ symmetry is broken spontaneously
to $SU(N)$ by the vacuum expectation value $\langle 0| U |0\rangle = \unitmatrix$, where $\unitmatrix$ is the $N\times N$ unit matrix.
This leads to the appearance of $N^2-1$ Goldstone bosons, which
correspond to the degrees of freedom contained in the matrix field $U$.
The term proportional to $M^2$ breaks the symmetry explicitly and causes
the Goldstone bosons to pick up a mass which,
at tree level, is equal to $M$. In terms of equal quark masses $\hat m$ we have
$M^2 = 2 B \hat m$.

The Lagrangian (\ref{lagrangian}) coincides with ChPT and therefore
constitutes an effective Lagrangian for two- and
three-flavour QCD for $N=2$ and $N=3$, respectively. Note, however, that
in the case considered here, all mesons have the same mass.
How this corresponds to a theory formulated in terms of quarks can be found in more detail in,
e.g., \cite{Bijnens:2009qm}. Below we occasionally use a vector notation for
quarks $q$ with $q^T= (q_1,\ldots,q_N)$ where the subscript denotes the flavour.

In previous publications~\cite{Bijnens:2009zi,Bijnens:2010xg,Bijnens:2012hf},
the chiral logarithms of
the massive nonlinear $O(n+1)/O(n)$ model have been considered. The two models
coincide for $N=2$ and $n=3$, such that the
corresponding results can be used as a check.

There are many ways the special unitary matrix $U$ can be parametrized in
terms of the meson matrix $\phi = \phi^a T^a$, where $T^a$ are the generators
of $SU(N)$ normalized as $\langle T^a T^b\rangle = \delta^{ab}$.
Physical results are independent of this choice.  As in the earlier work 
on the massive $O(n)$ model, one can therefore use different
parametrizations to obtain a thorough check of the calculation.
The four parametrizations we have used are
\begin{align} \label{params}
U_1 &= \exp\left(\frac{i\sqrt2}F\phi\right) \,, \qquad
&U_2 &= \frac{\unitmatrix+i\left(\beta_2+\frac{1}{\sqrt2 F}\phi\right)}
             {\unitmatrix-i\left(\beta_2+\frac{1}{\sqrt2 F}\phi\right)} \; , 
\nonumber\\[3mm]
U_3 &= e^{i\beta_3} \frac{\unitmatrix+\frac{i}{\sqrt2 F}\phi}
                        {\unitmatrix-\frac{i}{\sqrt2 F}\phi} \,, \qquad
&U_4 &= e^{i\beta_4}
\left(\sqrt{\unitmatrix-\frac{2}{F^2}\phi^2}+i\frac{\sqrt2}{F}\phi\right) \; .
\end{align}
The matrices must be special, i.e., $\det U_i=1$ which for $U_1$ is an automatic
consequence of $\langle \phi \rangle = 0$. For the other cases one has
to solve for $\beta_i$ in terms of $\phi$. Using $\langle \log U_i \rangle = \log\det U_i = 0$,
one finds that the $\beta_i$ start at order $\phi^3$.
For $\beta_2$ we can then solve the resulting equation perturbatively while
$\beta_3$ and $\beta_4$ can be written explicitly in terms of $\phi$ as
\begin{align}
\beta_3 &= \frac{i}{N}\left\langle\log\left(\unitmatrix+\frac{i\phi}{\sqrt2 F}\right)-\log\left(\unitmatrix-\frac{i\phi}{\sqrt2 F}\right)\right\rangle
= -\frac{2}{N}\sum_{n=1}^\infty\frac{(-1)^n}{2n+1}
\left\langle\left(\frac{\phi}{\sqrt{2}F}\right)^{2n+1}\right\rangle,
\nonumber\\[2mm]
\beta_4 &= -\frac{1}{N}\left\langle\arcsin\left(\frac{\sqrt{2}\phi}{F}\right) 
\right\rangle
= -\frac{1}{N}\sum_{n=1}^\infty\frac{(2n)!}{4^n(n!)^2(2n+1)}
\left\langle\left(\frac{\sqrt{2}\phi}{F}\right)^{2n+1}\right\rangle.
\end{align}
Note that the $n=0$ term vanishes in both sums since $\langle\phi\rangle=0$.
We could have added a fifth parametrization by adding a singlet component
to $\phi$ in $U_4$ as was done for parametrization~2.

It is also possible to treat $U(N)\times U(N)/U(N)$ by simply allowing
$\phi$ to have a singlet component and removing the $\beta_i$. We do not
discuss this case.

\section{Leading logarithms}
\label{sec:LLeven}

The method used here is entirely analogous to the work in
\cite{Bijnens:2009zi,Bijnens:2010xg,Bijnens:2012hf} but with $\phi$
a traceless $N\times N$ Hermitian matrix instead of an $n$-dimensional vector.
The calculations are done schematically as follows:
first we generate all needed one-loop diagrams
with a {\sc C++} program. The diagrams at each order are
then evaluated using {\sc FORM}~\cite{Vermaseren:2000nd}.
The integrals are performed using a recursive method.
The results are then combined to provide the needed Lagrangians at the next order.
A more detailed discussion of the method and the underlying principles can
be found in \cite{Bijnens:2009zi,Bijnens:2010xg}.

Flavour sums in the earlier work were rather trivial to
perform. Here, keeping track of the different terms is somewhat more
tricky but all the flavour sums can be performed using the methods
of \cite{Bijnens:2009qm}. The underlying idea is to use
\begin{align}
\left\langle T^a A T^a B\right\rangle = \left\langle A\right\rangle
\left\langle B\right\rangle
 -\frac{1}{N}\left\langle AB\right\rangle\,, \qquad
\left\langle T^a A\right\rangle\left\langle T^a B\right\rangle =
 \left\langle AB\right\rangle
 -\frac{1}{N}\left\langle A\right\rangle\left\langle B\right\rangle\,,
\end{align}
for the sums over the generators $T^a$.

In the following, we will present our results for the coefficients of the
leading logarithm contribution to several
physical quantities. In all cases, we have the choice of expressing
these in terms of lowest order or physical
parameters, which can have quite a substantial effect on the convergence
of the series. Following the definitions
in~\cite{Bijnens:2010xg} we expand a given observable $O_\phys$ as
\begin{eqnarray}
\label{defexpL}
O_\phys &=&  O_0 \left(1+a_1 L +a_2 L^2+\cdots\right)\,,
\\
\label{defexpLphys}
O_\phys &=&  O_0 \left(1+c_1 L_\phys +c_2 L_\phys^2+\cdots\right)\,,
\end{eqnarray}
where the chiral logarithms are defined either from the lowest-order parameters
$M$ and $F$ as
\begin{equation}
\label{defL}
 L \equiv \frac{M^2}{16\pi^2F^2} \log \frac{\mu^2}{M^2} \,,
\end{equation}
or from the physical mass $M_\pi$ and decay constant $F_\pi$ as
\begin{equation}
\label{defLphys}
 L_\phys \equiv \frac{M_\pi^2}{16\pi^2F_\pi^2} \log \frac{\mu^2}{M_\pi^2}\,.
\end{equation}
These are relevant for the static quantities where the mass is the only
dimensionful parameter. In general the argument of the logarithm is not uniquely
determined at the level of leading logarithms. For the cases with more
dimensionful quantities we usually use the more general
\begin{equation}
\label{FFlog}
 L_\M \equiv \frac{M_\pi^2}{16 \pi^2 F_\pi^2} \log \frac{\mu^2}{\M^2} \,,
\end{equation}
where $\M$ is some combination of the relevant dimensionful quantities.

\begin{table}[p]
\small
\begin{center}
\begin{tabular}{|c|c|c|l|}
\hline
$i$ & $a_i$ for $N=2$ & $a_i$ for $N=3$ & $a_i$ for general $N$\\
\hline
1 & $-1/2$ & $- 1/3 $        & $ - N^{-1} $\\[1mm]
2 & 17/8 & $27/8$       & $9/2\,N^{-2} - 1/2
          + 3/8\,N^2 $\\[1mm]
3 & $-103/24$ & $- 3799/648$    & $-89/3\,N^{-3}
          + 19/3\,N^{-1}
          - 37/24\,N
          - 1/12\,N^3 $ \\[1mm]
4 & 24367/1152 & $146657/2592$ & $2015/8\,N^{-4}
          - 773/12\,N^{-2}
          + 193/18
          + 121/288\,N^2$\\
& & &$
          + 41/72\,N^4$ \\[1mm]
5 & $-8821/144$ & $- 27470059/186624$   & $- 38684/15\,N^{-5}
          + 6633/10\,N^{-3}
          - 59303/1080\,N^{-1}$\\
& & &$
          - 5077/1440\,N
          - 11327/4320\,N^3
          - 8743/34560\,N^5 $\\[1mm]
6$^*$ & $\frac{1922964667}{6220800}$ & $\frac{12902773163}{9331200}$ & $7329919/240\,N^{-6} - 1652293/240\,N^{-4}$\\
& & &$    - 4910303/15552\,N^{-2}
          + 205365409/972000$\\
& & &$
          - 69368761/7776000\,N^2
          + 14222209/2592000\,N^4$\\
& & &$
          + 3778133/3110400\,N^6
$\\
\hline
\end{tabular}
\end{center}
\caption{\label{tabmass1}The coefficients $a_i$ of the leading logarithm $L^i$
up to $i=6$ for the physical meson mass for the physical cases $N=2$ and
$N=3$ as well as for general $N$.}
\end{table}
\begin{table}[p]
\small
\begin{center}
\begin{tabular}{|c|c|c|l|}
\hline
$i$ & $c_i$ for $N=2$ & $c_i$ for $N=3$ & $c_i$ for general $N$\\
\hline
1 & $ - 1/2$ & $-1/3 $ & $ - N^{-1} $\\[1mm]
2 & $ 7/8 $  & $163/72 $ & $7/2\,N^{-2} - 3/2 + 3/8\,N^2  $\\[1mm]
3 & $ 211/48 $ & $9329/648  $ & $- 109/6\,N^{-3} + 71/6\,N^{-1} - 55/24\,N +
2/3\,N^3 $\\[1mm]
4 & $ 21547/1152 $ & $638861/7776$ & $3185/24\,N^{-4} - 1067/12\,N^{-2} + 1795/72
$\\
 & & & $
- 677/288\,N^2 + 77/72\,N^4 $\\[1mm]
5 & $ 179341/2304 $ & $\frac{86965829}{186624} $ &$- 150877/120\,N^{-5} +
47473/60\,N^{-3} - 23407/120\,N^{-1} $\\
 & & & $
 + 41713/1440\,N +
      4891/8640\,N^3 + 57557/34560\,N^5
 $\\[1mm]
6$^*$ & $ \frac{2086024177}{6220800} $ & $\frac{38355806767}{13996800} $ & $
229179/16\,N^{-6} - 392659/48\,N^{-4} + 106873049/77760\,N^{-2}
$\\
 & & & $+ 2800699/60750
       - 277103161/7776000\,N^2 $\\
 & & & $+ 68361593/5184000\,N^4 + 8001833/3110400\,N^6  $\\
\hline
\end{tabular}
\end{center}
\caption{\label{tabmass2} The coefficients $c_i$ of the leading logarithm $L^i_\phys$ up to $i=6$ for the
physical meson mass for the physical cases $N=2$ and $N=3$ as well as for general $N$.}
\end{table}

\section{Mass}
\label{secmass}
The mass is known fully to one \cite{Gasser:1986vb}
and two loops \cite{Bijnens:2009qm}. We have calculated the leading logarithms
to six-loop order here. For this we have computed the generic two-point
function of the $\phi$ fields in all four parametrizations and extracted
the mass as well as the wave function renormalization, which will be needed later.
The result is expressed in the form of ~(\ref{defexpL}) and~(\ref{defexpLphys})
with $O_\phys = M_\pi^2$ and $O_0=M^2$.  The first six coefficients $a_i$ and $c_i$
of the expansions of the physical mass are listed in
Tables~\ref{tabmass1} and~\ref{tabmass2}.
The coefficients for $N=2$ agree with the results
from~\cite{Bijnens:2009zi,Bijnens:2010xg,Bijnens:2012hf} and with the one- and two-loop
results from~\cite{Gasser:1986vb,Bijnens:2009qm}.

Since the calculation is very time consuming,
the sixth order has only been checked with two of the four parametrizations in \eqref{params}.
Throughout the  paper, results with this limitation are marked by an asterisk next to
the number that labels the order.

It is rather clear from the expressions that there is a pattern in the
powers of $N$ that appear. They always jump by powers of 2. Similar steps
can be seen in all results quoted in this paper. This is due to the
$SU(N)$ group structure of all flavour traces that need to be evaluated as is
proven in general in Appendix~\ref{appA}.

\begin{figure}[p]
\begin{minipage}{0.49\textwidth}
\includegraphics[width=\textwidth]{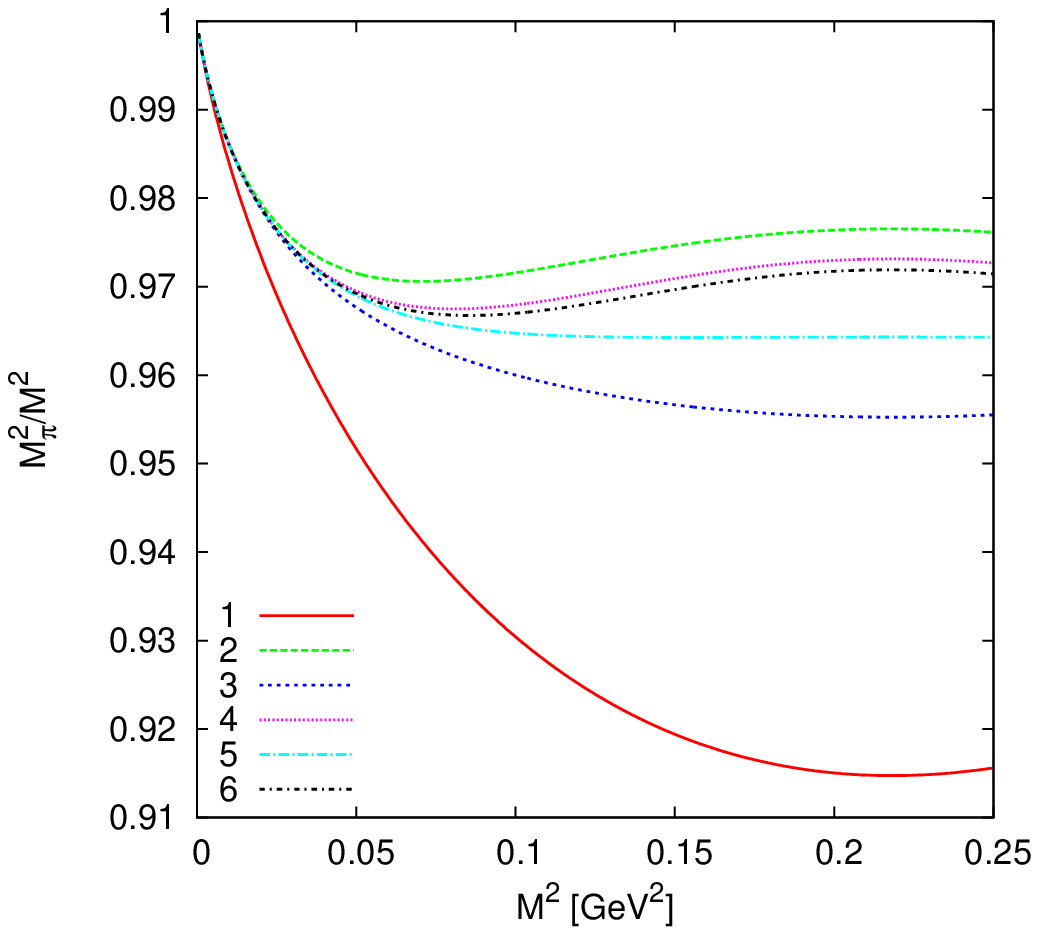}
\end{minipage}
\begin{minipage}{0.49\textwidth}
\includegraphics[width=\textwidth]{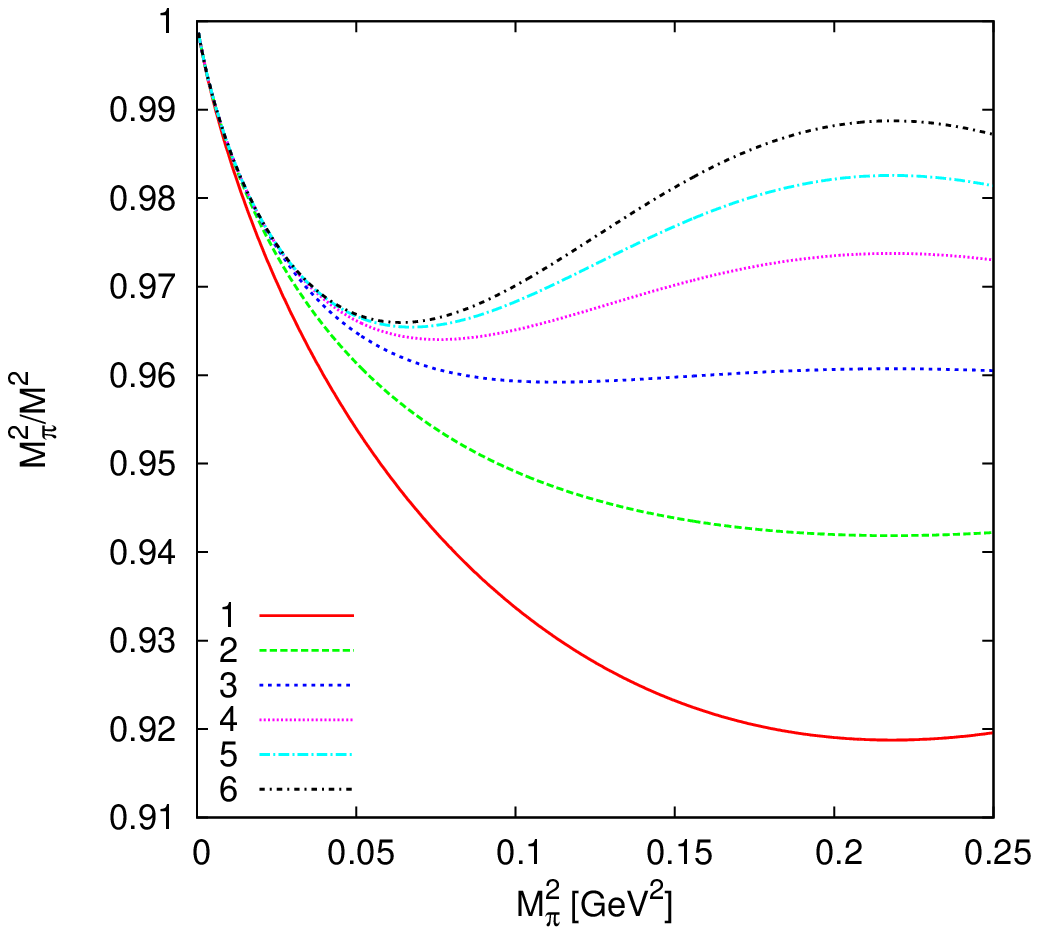}
\end{minipage}
\caption{\label{figmass1}
The contribution of the leading logarithms to $M^2_\pi/M^2$ order by order for
$F=0.090$~GeV, $F_\pi=0.0922$~GeV, $\mu =0.77$~GeV and $N=2$.
The left panel shows the expansion in $L$ keeping $F$ fixed,
the right panel the expansion in $L_{\phys}$ keeping $F_\pi$ fixed.
Plots similar to Fig.~1 in \cite{Bijnens:2012hf}.}
\end{figure}

\begin{figure}[p]
\begin{minipage}{0.49\textwidth}
\includegraphics[width=\textwidth]{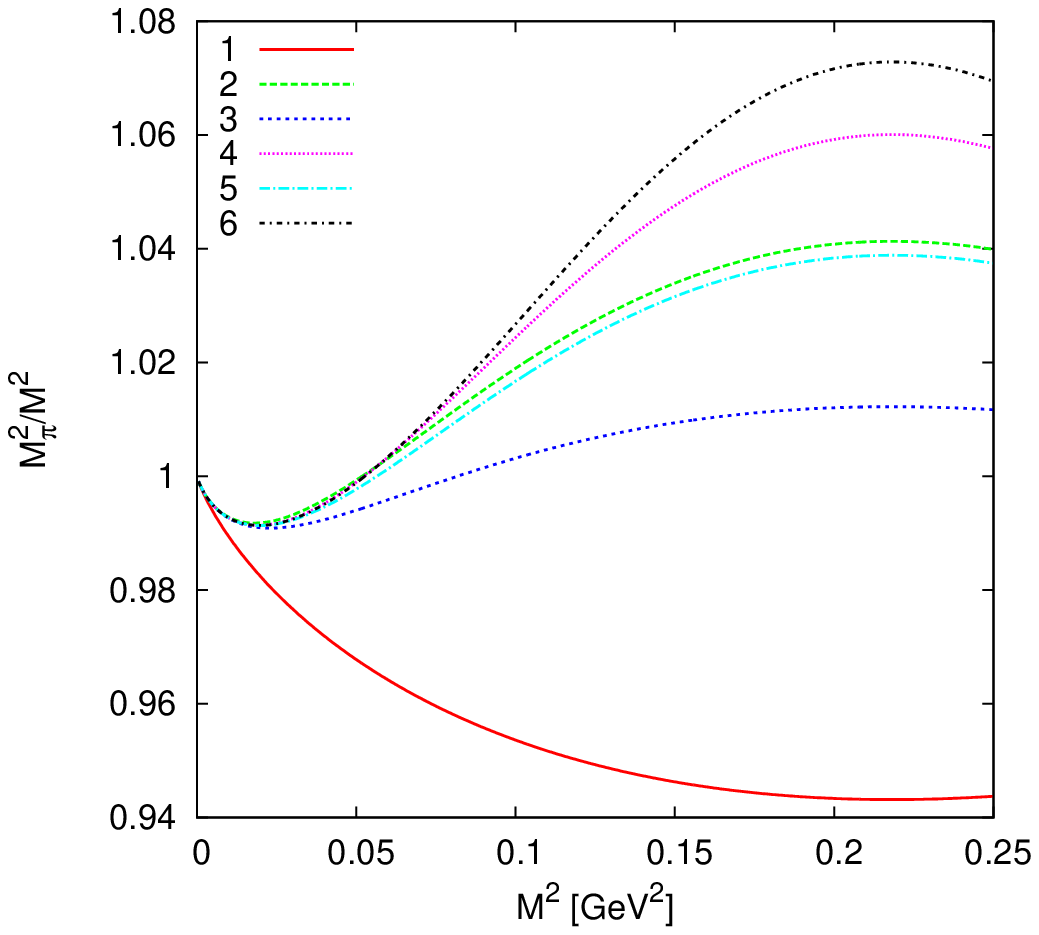}
\end{minipage}
\begin{minipage}{0.49\textwidth}
\includegraphics[width=\textwidth]{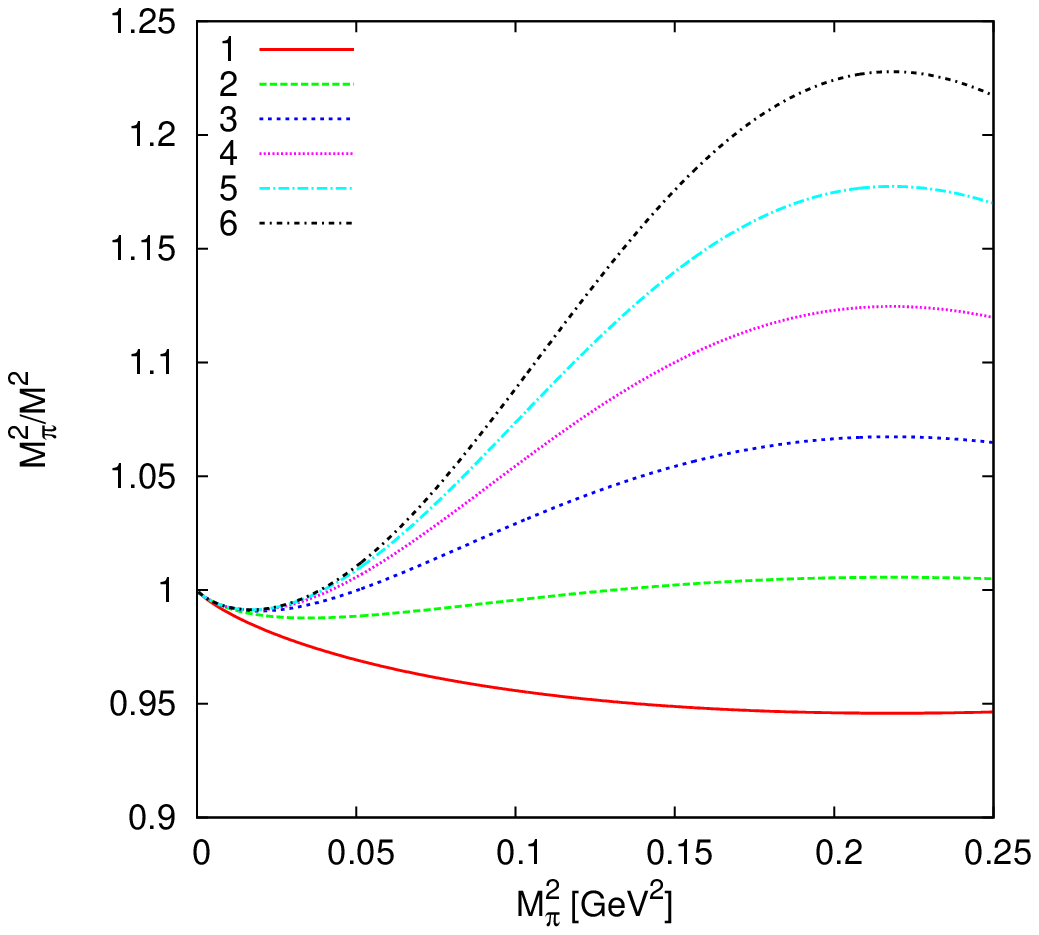}
\end{minipage}
\caption{\label{figmass2}
The contribution of the leading logarithms to $M^2_\pi/M^2$ order by order for
$F=0.090$~GeV, $F_\pi=0.0922$~GeV, $\mu =0.77$~GeV and $N=3$.
The left panel shows the expansion in $L$ keeping $F$ fixed,
the right panel the expansion in $L_{\phys}$ keeping $F_\pi$ fixed.}
\end{figure}

We can use our results to check the convergence of the two expansions.
In Fig.~\ref{figmass1} for $N=2$ and Fig.~\ref{figmass2} for $N=3$,
the input values chosen are $F=0.090$~GeV for the expansion in terms of $L$
and $F_\pi = 0.0922$~GeV for the expansion in $L_\phys$ as well as
$\mu=0.77$~GeV. The convergence is somewhat worse for $N=3$ than for $N=2$.

\section{Decay constant}
\label{secdecay}

The decay constant $F_\pi$ is defined by
\begin{equation}
\langle 0| j^b_{A,\mu}| \phi^a(p)\rangle = i\sqrt{2} F_\pi
p_\mu\delta^{ab}
\end{equation}
for a meson corresponding to the quark flavour combination
$i \bar q \gamma_5 T^a q$ and the axial current $\bar q \gamma_\mu \gamma_5 T^b q$.
Note that $F_\pi$ is equal for all mesons since we are in the equal mass limit.
The decay constant is known fully to one \cite{Gasser:1986vb}
and two loops \cite{Bijnens:2009qm}. Here, we evaluate the leading logarithms
to six loops.

We need to evaluate a matrix-element with one external axial current
and one incoming meson. The diagrams required for the wave function
renormalization were already done in the mass calculation 
in the previous section. We thus need to evaluate all
relevant one-particle-irreducible (1PI) diagrams with an external $a^a_\mu$. 
Up to the more complicated group theory the calculation is the same as
in our earlier work.

We give the first six coefficients for both leading logarithm series with $O_\phys = F_\pi$
and $O_0 = F$ in Tables~\ref{tab:decay} and~\ref{tab:decay2}. 
Note that once $F_\pi$ is known as a function of
$F$, we can express all observables as a function of the
physical $M^2_\pi$ and $F_\pi$. We already used this to get the
coefficients $c_i$ in Tables~\ref{tabmass2} and~\ref{tab:decay2}
from the corresponding $a_i$.

\begin{table}[t]
\small
\begin{center}
\begin{tabular}{|c|c|c|l|}
\hline
$i$ & $a_i$ for $N=2$ & $a_i$ for $N=3$ & $a_i$ for general $N$\\
\hline
1 & $ 1$ & $3/2 $ & $
1/2\,N
          $\\[1mm]
2 & $ - 5/4$ & $-35/16 $ &  $
- 1/2 - 3/16\,N^2
          $\\[1mm]
3 & $ 83/24 $ & $293/36 $ & $
23/12\,N^{-1} + 1/4\,N + 1/4\,N^3
          $\\[1mm]
4 & $ - 3013/288 $ & $-\frac{413359}{13824} $ & $
- 139/12\,N^{-2} + 7/54 - 523/576\,N^2 - 3511/13824\,N^4
          $\\[1mm]
5 & $  \frac{2060147}{51840} $ & $\frac{96197471}{622080} $ & $
22357/240\,N^{-3} - 5063/648\,N^{-1} + 16157/5184\,N$\\
& & & $
          + 26237/17280\,N^3
          + 5885/13824\,N^5
          $\\[1mm]
6$^*$ & $ -\frac{69228787}{466560}  $ & $-\frac{932532830269}{1343692800} $ & $
          - 41296/45\,N^{-4}
          + 5690093/58320\,N^{-2}
$\\ & & & $
					+ 14622631/9331200
					- 1944182341/186624000\,N^2
$\\ & & & $
					- 945730747/373248000\,N^4
					- 81119291/149299200\,N^6
$\\
\hline
\end{tabular}
\end{center}
\caption{\label{tab:decay}The coefficients $a_i$ of the leading logarithm $L^i$ up to $i=6$ for the
decay constant $F_\pi$ for the physical cases $N=2$ and $N=3$ as well
as for general $N$.}
\end{table}
\begin{table}[t]
\small
\begin{center}
\begin{tabular}{|c|c|c|l|}
\hline
$i$ & $c_i$ for $N=2$ & $c_i$ for $N=3$ & $c_i$ for general $N$\\
\hline
1 & $ 1 $ & $ 3/2 $ & $ 1/2\,N $\\[1mm]
2 & $ 5/4 $ & $45/16 $ & $5/16\,N^2$\\[1mm]
3 & $  13/12 $ & $131/36 $ & $-1/3\,N^{-1} + 1/8\,N + 1/8\,N^3$\\[1mm]
4 & $  - 577/288$ & $-\frac{113471}{13824} $ & $9/4\,N^{-2} + 209/576\,N^2
- 229/108 - 1639/13824\,N^4 $\\[1mm]
5 & $   - 14137/810 $ & $-\frac{65712649}{622080}$ & $- 1097/60\,N^{-3} +
11095/648\,N^{-1} - 40225/5184\,N $\\ & & & $
+ 2137/2880\,N^3 - 679/
      1536 N^5      
$\\[1mm]
6$^*$ & $-\frac{37737751}{466560} $ & $-\frac{889506647989}{1343692800}$  &
$6745/36\,N^{-4} - 9274909/58320\,N^{-2} + 611736991/9331200
$\\ & & & $
- 3858946741/
      186624000 N^2 + 440983853/373248000\,N^4
$\\ & & & $
 - 127342211/149299200\,N^6
$\\
\hline
\end{tabular}
\end{center}
\caption{\label{tab:decay2}The coefficients $c_i$ of the leading
logarithm $L^i_\phys$ up to $i=6$ for the
decay constant $F_\pi$ for the physical cases $N=2$ and $N=3$ as well
as for general $N$.}
\end{table}

\begin{figure}[tp]
\begin{minipage}{0.49\textwidth}
\includegraphics[width=\textwidth]{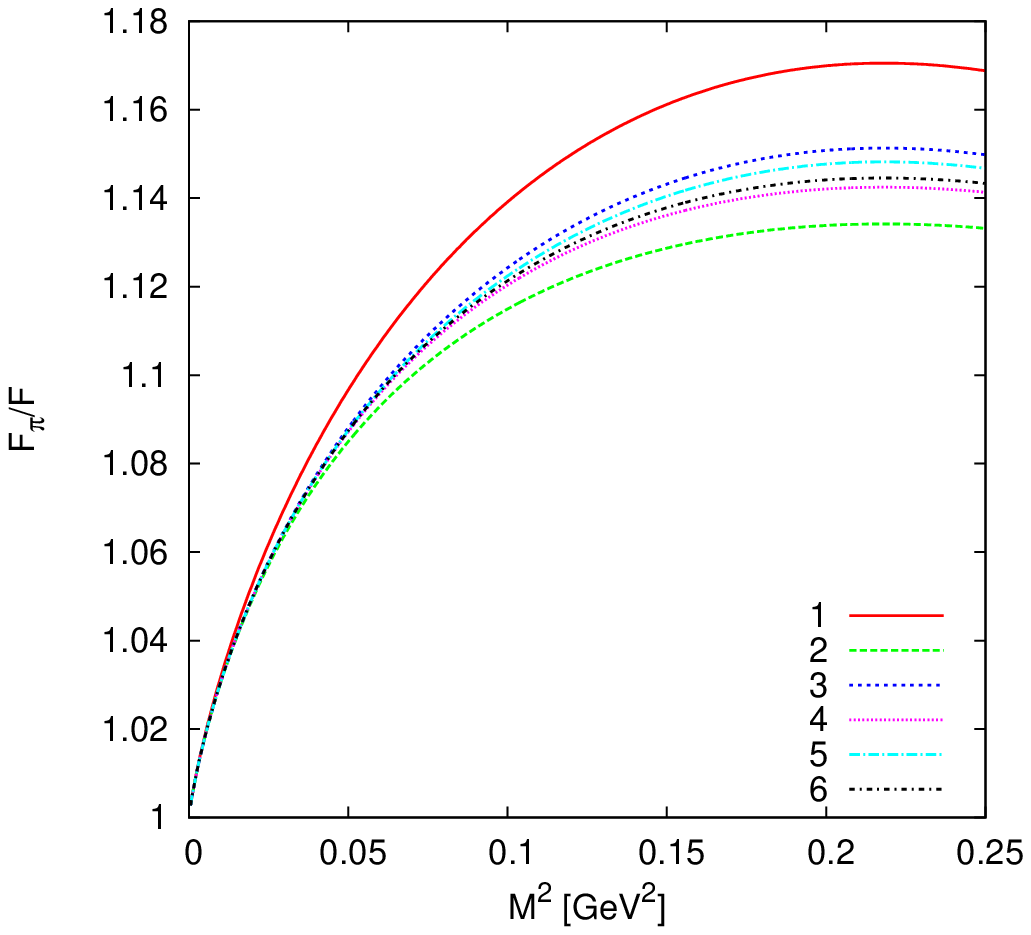}
\end{minipage}
\begin{minipage}{0.49\textwidth}
\includegraphics[width=\textwidth]{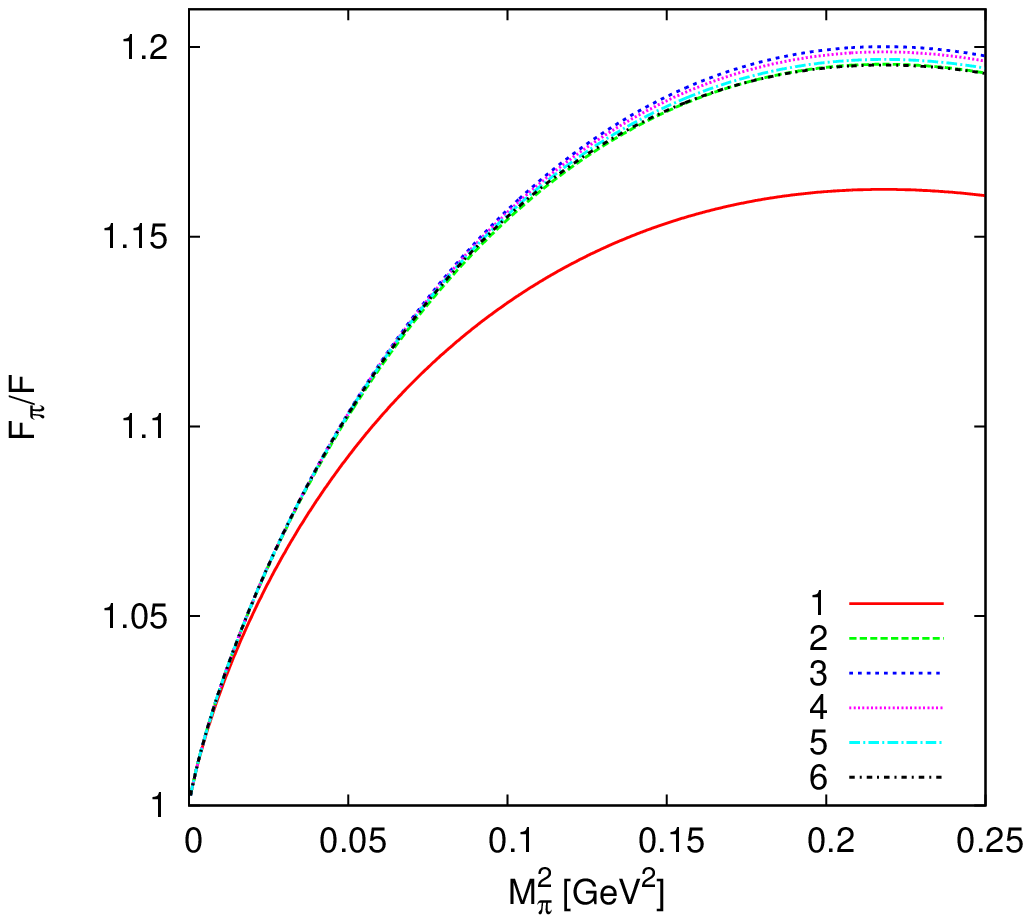}
\end{minipage}
\caption{\label{figdecay1}
The contribution of the leading logarithms to $F_\pi/F$ order by order for
$F=0.090$~GeV, $F_\pi=0.0922$~GeV, $\mu =0.77$~GeV and $N=2$.
The left panel shows the expansion in $L$ keeping $F$ fixed,
the right panel the expansion in $L_{\phys}$ keeping $F_\pi$ fixed.
Plots similar to Fig.~2 in \cite{Bijnens:2012hf}.}
\end{figure}

\begin{figure}[tp]
\begin{minipage}{0.49\textwidth}
\includegraphics[width=\textwidth]{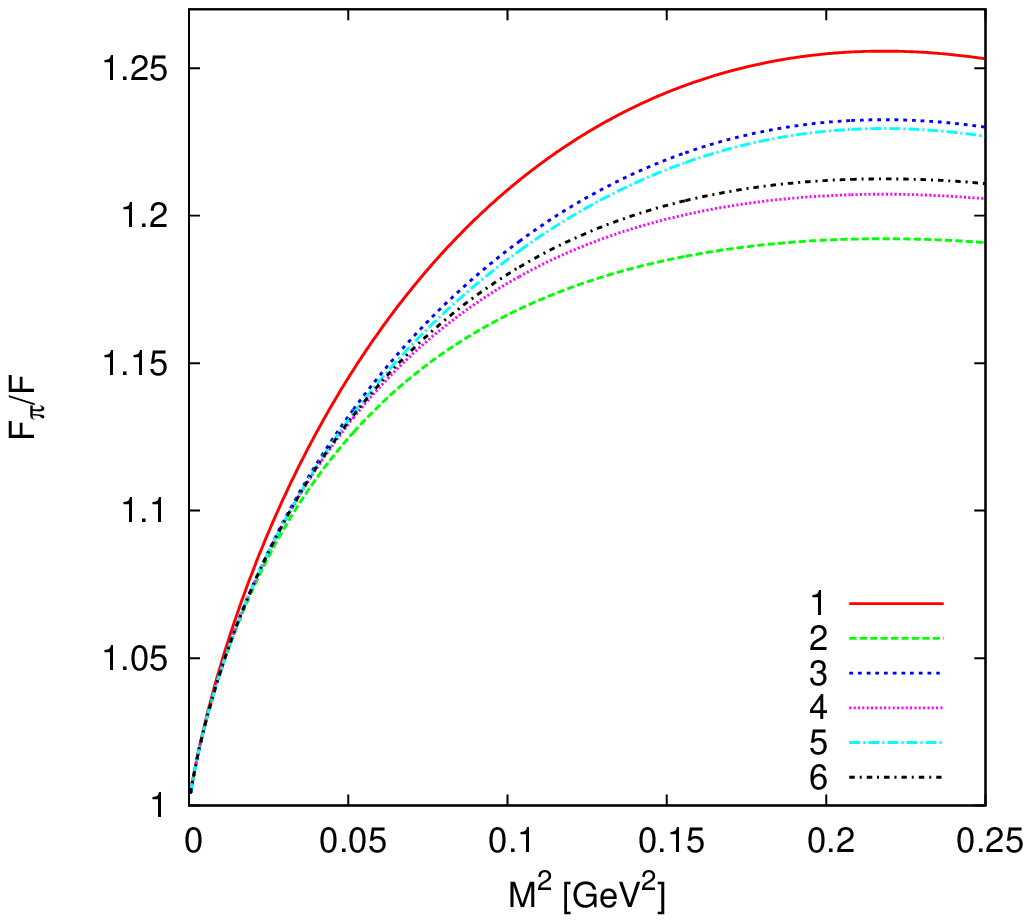}
\end{minipage}
\begin{minipage}{0.49\textwidth}
\includegraphics[width=\textwidth]{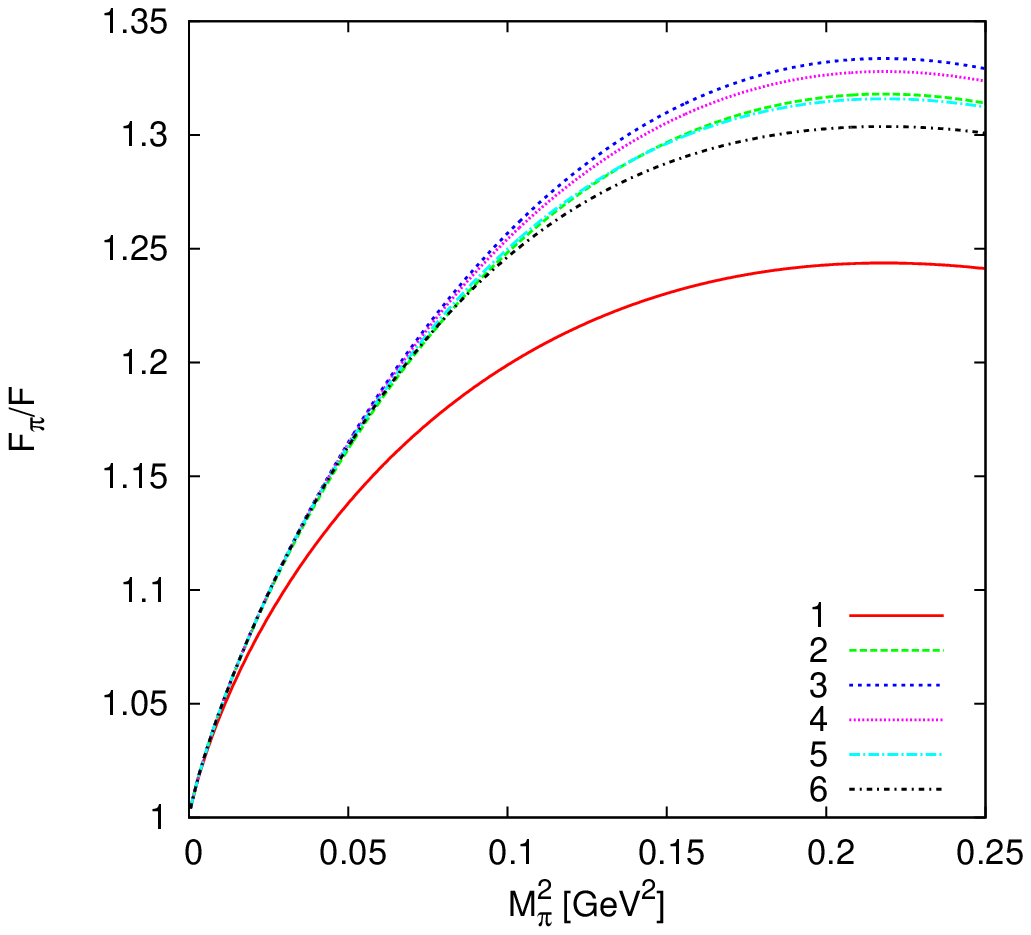}
\end{minipage}
\caption{\label{figdecay2}
The contribution of the leading logarithms to $F_\pi/F$ order by order for
$F=0.090$~GeV, $F_\pi=0.0922$~GeV, $\mu =0.77$~GeV and $N=3$.
The left panel shows the expansion in $L$ keeping $F$ fixed,
the right panel the expansion in $L_{\phys}$ keeping $F_\pi$ fixed.}
\end{figure}

We have plotted in Figs.~\ref{figdecay1} and \ref{figdecay2}
the expansion in terms of the unrenormalized quantities and in terms of
the physical quantities for $N=2$ and $N=3$ respectively.
In both cases we get convergence but it is better for the expansion in
physical quantities. It is also much better for $N=2$ than for $N=3$.


\section{Vacuum expectation value}
\label{vev}

The expression for the leading logarithms of the vacuum expectation
value (VEV) follows from the definition
\begin{align}
V_\phys = \langle 0 | -j_0^s | 0 \rangle \,,
\end{align}
where $j_0^s$ is the QCD current associated with the
scalar external source $s$ introduced
in~(\ref{chiDef}) with the singlet generator normalized to 1.
In terms of quarks the definition is
\begin{align}
\langle 0 | \bar q_i q_j | 0 \rangle = V_\phys \delta_{ij}\,.
\end{align}
At lowest order, $V_\phys \equiv V_0 = -2 B F^2$.
The VEV is known fully to one \cite{Gasser:1986vb}
and two loops \cite{Bijnens:2009qm}. Here we evaluate the leading logarithms
to seven loops.

The first seven coefficients of the expansions in (\ref{defexpL})
and (\ref{defexpLphys}) for $O_\phys = V_\phys$ and $O_0 = V_0$
are given in Tables~\ref{tab:vev} and~\ref{tab:vev2}, respectively.

\begin{figure}[p]
\begin{minipage}{0.49\textwidth}
\includegraphics[width=\textwidth]{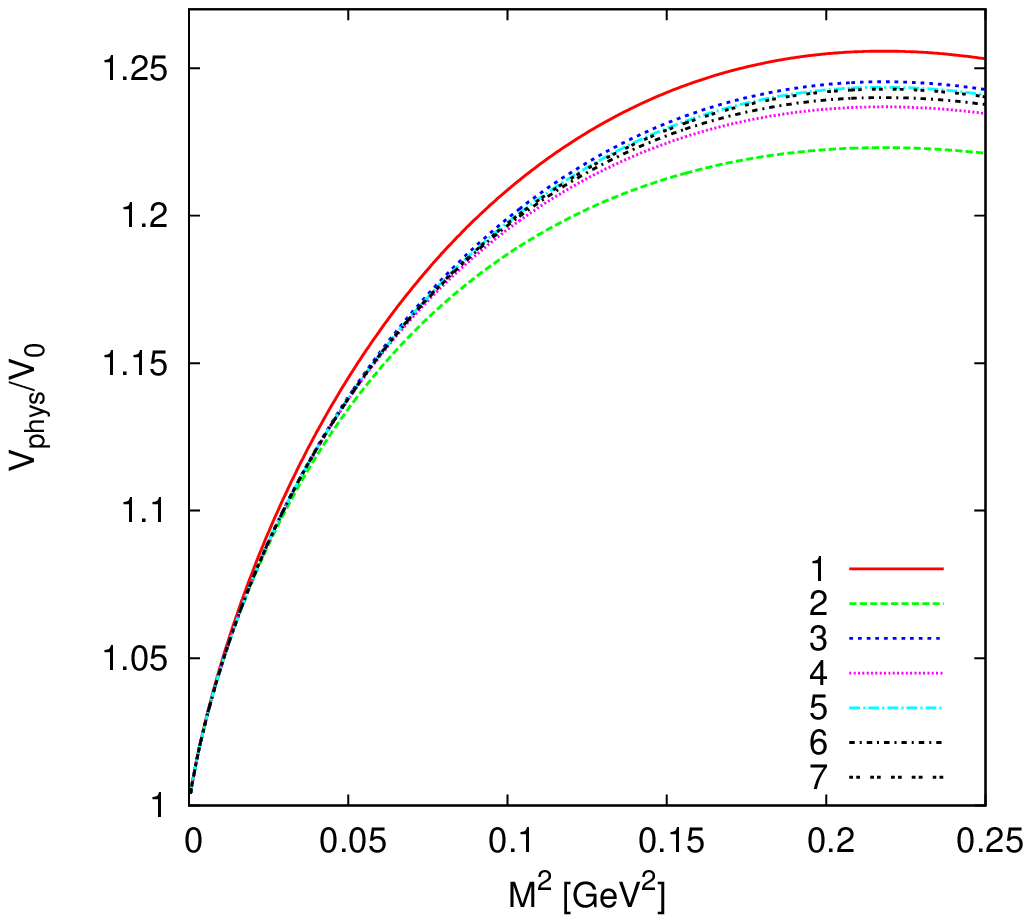}
\end{minipage}
\begin{minipage}{0.49\textwidth}
\includegraphics[width=\textwidth]{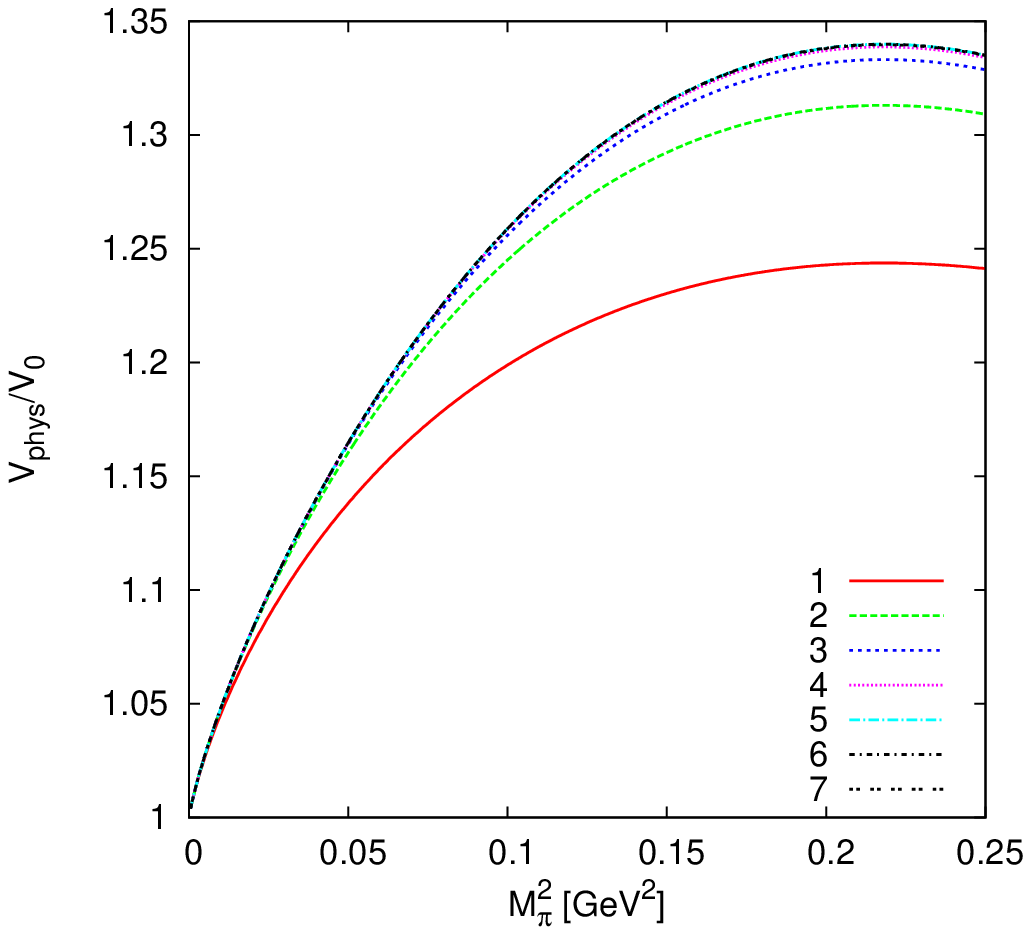}
\end{minipage}
\caption{\label{figvev1}
The contribution of the leading logarithms to $V_\phys/V_0$ order by order for
$F=0.090$~GeV, $F_\pi=0.0922$~GeV, $\mu =0.77$~GeV and $N=2$.
The left panel shows the expansion in $L$ keeping $F$ fixed,
the right panel the expansion in $L_{\phys}$ keeping $F_\pi$ fixed.}
\end{figure}

\begin{figure}[p]
\begin{minipage}{0.49\textwidth}
\includegraphics[width=\textwidth]{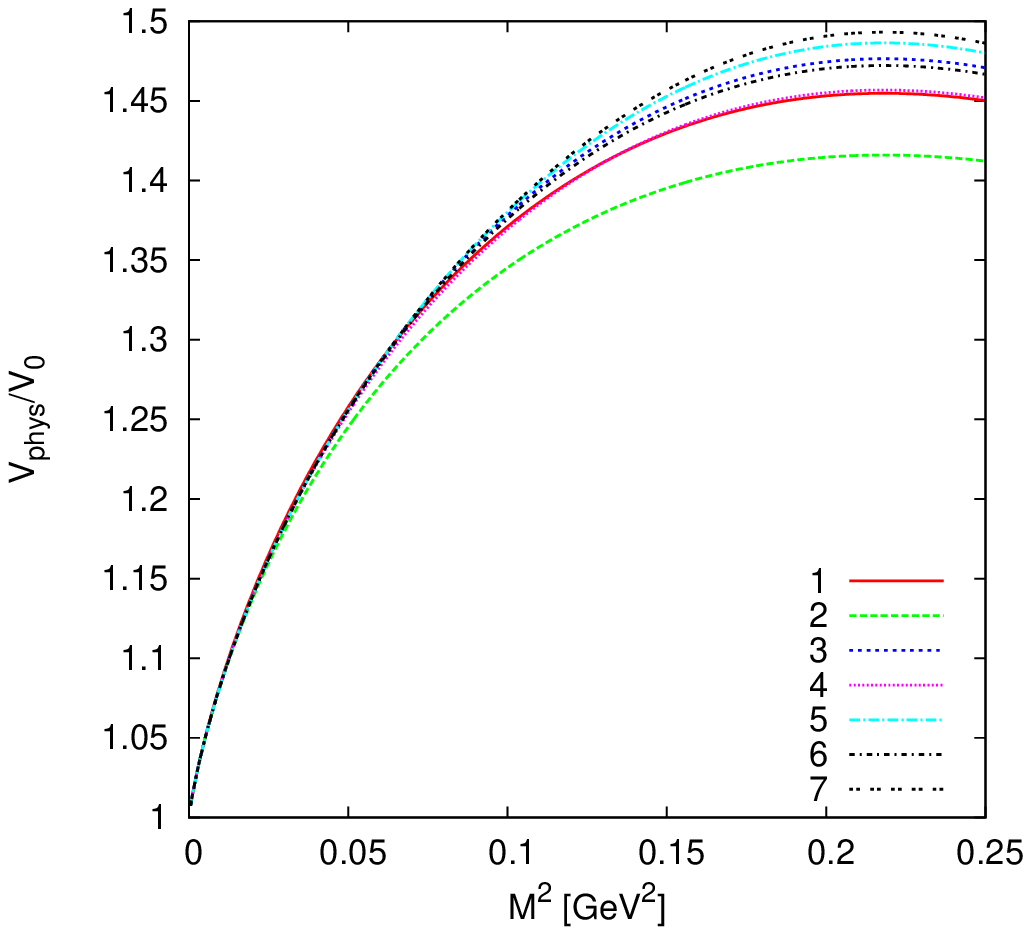}
\end{minipage}
\begin{minipage}{0.49\textwidth}
\includegraphics[width=\textwidth]{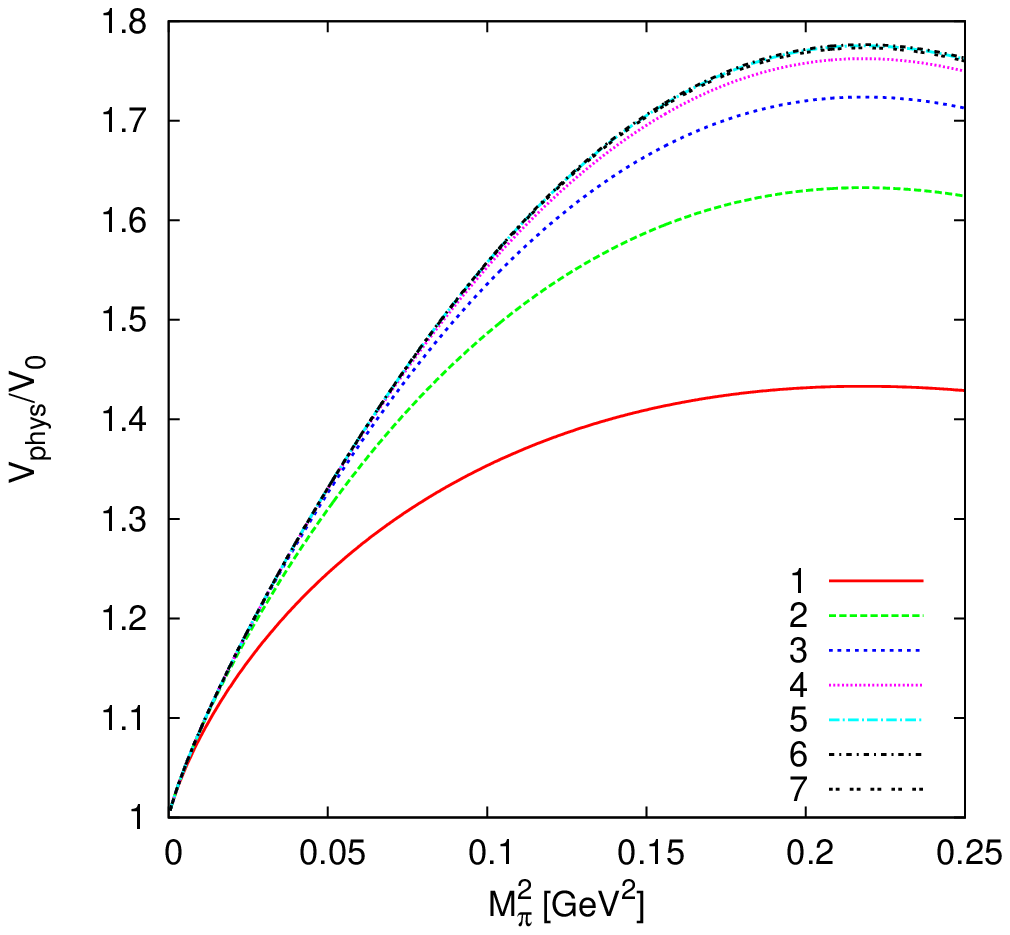}
\end{minipage}
\caption{\label{figvev2}
The contribution of the leading logarithms to $V_\phys/V_0$ order by order for
$F=0.090$~GeV, $F_\pi=0.0922$~GeV, $\mu =0.77$~GeV and $N=3$.
The left panel shows the expansion in $L$ keeping $F$ fixed,
the right panel the expansion in $L_{\phys}$ keeping $F_\pi$ fixed.}
\end{figure}

We have plotted in Figures~\ref{figvev1} and \ref{figvev2} the expansion
in terms of the
unrenormalized quantities and in terms of the physical quantities for
$N=2$ and $N=3$, respectively.
In both cases we get a good convergence but it is excellent for the expansion in
physical quantities.

\begin{table}[p]
\small
\begin{center}
\begin{tabular}{|c|c|c|l|}
\hline
$i$ & $a_i$ for $N=2$ & $a_i$ for $N=3$ & $a_i$ for general $N$\\
\hline
1 &     3/2  & 8/3 &  $ - N^{-1} + N $\\
2 &   $- 9/8$  & $-4/3$ & $3/2\,N^{-2} - 3/2 $\\
3 &   $ 9/2$ & $988/81$ &  $- 20/3\,N^{-3} + 22/3\,N^{-1} - 7/6\,N + 1/2\,N^3 $  \\
4 &  $- 1285/128$ & $-5660/243 $ &  $1025/24\,N^{-4} - 205/4\,N^{-2} + 175/16 - 55/24\,N^2 - 5/48\,N^4$ \\
5 &  $ 46$ & $\frac{399563}{1944}$ &  $- 350 N^{-5} + 2188/5\,N^{-3} - 12539/120\,N^{-1} + 1321/80\,N$\\
 & & & $ - 373/960\,N^3 + 737/960\,N^5$\\
6$^*$ & $-\frac{1305605}{9216}$ & $-\frac{242777185}{419904}$ & $2490019/720\,N^{-6} - 3137701/720\,N^{-4}$ \\
& & & $+ 12971623/12960\,N^{-2} - 1295581/12960$\\
& & & $+ 154399/51840\,N^2 - 277697/69120\,N^4 - 68761/207360\,N^6$\\
7$^*$ & $\frac{153149887}{226800}$ & $\frac{137725650367}{27556200}$
& $-12489752/315\,N^{-7} + 15424312/315\,N^{-5}$\\
& & & $
 - 79037542/
 8505 N^{-3} - 129606331/567000\,N^{-1}$\\
& & & $ + 66338023/324000\,N -
 19464419/3402000\,N^3 $\\
& & & $+ 67022189/13608000\,N^5 +
 4453133/2721600\,N^7$ \\
\hline
\end{tabular}
\end{center}
\caption{\label{tab:vev} The coefficients $a_i$ of the leading logarithm
$L^i$ up to $i=7$ for the vacuum expectation value $V_\phys$ for the physical
cases $N=2$ and $N=3$ as well as for general $N$.}
\end{table}
\begin{table}[p]
\small
\begin{center}
\begin{tabular}{|c|c|c|l|}
\hline
$i$ & $c_i$ for $N=2$ & $c_i$ for $N=3$ & $c_i$ for general $N$\\
\hline
1 &  3/2 &  8/3  & $- N^{-1} + N $ \\
2 &  21/8 & 68/9   & $1/2\,N^{-2} - 3/2 + N^2 $ \\
3 & 75/16  & 1720/81  & $ - 7/6\,N^{-3} + 7/3\,N^{-1} - 13/6\,N + N^3$ \\
4 & 1023/128 & 26881/486 & $109/24\,N^{-4} - 103/12\,N^{-2} + 277/48$ \\
& & & $- 127/48\,N^2 + 11/12\,N^4 $\\
5 & 2669/256 & 82861/729  & $ - 637/24\,N^{-5} + 5587/120\,N^{-3} -
57887/2160\,N^{-1} $\\
  & &  &$ + 9241/1080\,N - 5263/
      2160 N^3 + 179/270\,N^5$ \\
6$^*$ & $-\frac{48029}{138240}$ & $\frac{67564919}{1049760}$ &
$150877/720\,N^{-6} - 49505/144\,N^{-4} + 46879/288\,N^{-2}$\\
 & & & $- 378373/12960 +
      9427/10368\,N^2 - 2741/6912\,N^4 $\\
 & & & $+ 14701/103680\,N^6$\\
7$^*$ & $-\frac{6924628769}{87091200}$ & $-\frac{966193799261}{881798400} $
& $-229179/112\,N^{-7} + 440981/140\,N^{-5}$\\
 & & & $ - 967456169/
 816480 N^{-3} - 986277601/163296000\,N^{-1}$\\
 & & & $ + 3916037663/31104000\,N -
 56820907057/1306368000\,N^3 $\\
 & & & $+ 7411227769/1306368000\,N^5 -
 19637251/26127360\,N^7$\\
\hline
\end{tabular}
\end{center}
\caption{\label{tab:vev2} The coefficients $c_i$ of the leading logarithm
$L^i_\phys$ up to $i=7$ for the vacuum expectation value $V_\phys$ for
the physical cases $N=2$ and $N=3$ as well as for general $N$.}
\end{table}


\section{Vector form factor} \label{secvff}

We turn now to the vector form factor which is defined by
\begin{equation}
\langle\phi^a(p_f)|
  j^{c}_{V,\mu}|\phi^{b}(p_i)\rangle
= \left\langle T^c\left(T^bT^a-T^aT^b\right)\right\rangle (p_f+p_i)_\mu
  F_V\left[(p_f-p_i)^2\right]\,.
\end{equation}
The vector current is $\bar q \gamma_\mu T^c q$.
It is known fully in two- and three-flavour ChPT to 
one \cite{Gasser:1983yg,Gasser:1984ux} and
two loops \cite{Bijnens:1998fm,Bijnens:2002hp}. Here we calculate the
leading logarithms in the equal mass case to five loops.

The procedure to find the leading logarithms is entirely the same as
in the earlier work \cite{Bijnens:2010xg,Bijnens:2012hf} with the modifications
needed for the more complicated flavour structure.
We express the result  in terms of $\tilde t= t/M_\pi^2$ and
the logarithm (\ref{FFlog})
with a scale $\M^2$ that is some combination of $t$ and $M_\pi^2$.
To fifth order we find
{\small
\begin{align}
\label{resultFV}
F_V&(t) = 1+ L_{\M} \Big[
           N/12\,\tilde t
          \Big]
       +  L_{\M}^2  \Big[
          \tilde t\,(1/12 + N^2/16)
          +\tilde t^2\,( N^2/288 )
           \Big]\nonumber \\[1mm]&
       +  L_{\M}^3   \Big[
         \tilde t \,(-1/12\,N^{-1} + 29/72\,N + 179/2592\,N^3 )
         + \tilde t^2\, (-137/648\,N-5/2592\,N^3)
\nonumber\\&
         + \tilde t^3\, ( N/32 + 5/5184\,N^3  )
          \Big]
\nonumber\\[1mm]
&       + L_{\M}^4   \Big[
          \tilde t \,( 17/18\,N^{-2} - 617/576 + 1111/768\,N^2 + 12011/124416\,N^4)
\nonumber\\&
         + \tilde t^2 \,(445/7776 - 79861/103680\,N^2 - 5519/414720\,N^4 )
\nonumber\\&
         + \tilde t^3 \,( 79459/777600\,N^2 + 407/172800\,N^4)
         +\tilde t^4\, (-N^2/345600 + N^4/518400 )
           \Big]
\nonumber\\[1mm]
&         + L_{\M}^5   \Big[
          \tilde t \,(- 391/45\,N^{-3} + 134797/12960\,N^{-1} - 5429753/933120\,N + 5232541/1244160\,N^3
\nonumber\\&
          + 14129/93312\,N^5 )
          + \tilde t^2 \,(-181271/233280\,N^{-1} + 194389/405000\,N
          \nonumber\\&
          - 262760809/116640000\,N^3 - 61724321/1866240000\,N^5)
           + \tilde t^3 \,(35/384\,N^{-1}
           \nonumber\\&
            + 1261489/23328000\,N + 35449669/90720000\,N^3 + 129189077/26127360000\,N^5)
\nonumber\\&
           + \tilde t^4 \,(-226531/4320000\,N - 58095211/1632960000\,N^3 - 935713/4354560000\,N^5)
\nonumber\\&
           + \tilde t^5 \,(871/115200\,N + 545009/181440000\,N^3 + 126059/6531840000\,N^5)
          \Big]
\,.
\end{align}}
Note that $F_V(0) =1$ as it should be.

\begin{table}[t]
\small
\begin{center}
\begin{tabular}{|c|c|c|l|}
\hline
$i$ & $c_i$ for $N=2$ & $c_i$ for $N=3$ & $c_i$ for general $N$\\
\hline
1  & 1  & 3/2         &  $N/2$ \\[1mm]
2  & 2  & 31/8        &   $ 1/2 + 3/8\,N^2 $\\[1mm]
3  & 853/108  & 877/48  &  $-1/2\,N^{-1} + 29/12\,N + 179/432\,N^3 $ \\[1mm]
4 & 50513/1296 & 824171/6912 &  $17/3\,N^{-2} - 617/96 + 1111/128\,N^2 + 12011/20736\,N^4$\\[1mm]
5  & 120401/648 & 33850135/41472 & $-782/15\,N^{-3} + 134797/2160\,N^{-1} - 5429753/155520\,N$\\
& & & $+ 5232541/207360\,N^3 + 14129/15552\,N^5$\\
\hline
\end{tabular}
\end{center}
\caption{\label{tab:radiusV}The coefficients $c_i$ of the leading logarithm $L^i_\phys$ up to $i=5$ for the
vector radius $\langle r^2 \rangle_V$ for the physical cases $N=2$ and
$N=3$ as well as for general $N$ in units of $M_\pi^2$.}
\end{table}

\begin{table}[t]
\small
\begin{center}
\begin{tabular}{|c|c|c|l|}
\hline
$i$ & $c_i$ for $N=2$ & $c_i$ for $N=3$ & $c_i$ for general $N$\\
\hline
1  & 0 & 0        &  0\\[1mm]
2  & $1/72$ &   1/32       &   $N^2/288 $\\[1mm]
3 & $-71/162$  & $-593/864$  &  $-137/648\,N -5/2592\,N^3 $
\\[1mm]
4 & $-25169/7776$ & $-1978981/248832$ &  $445/7776-79861/103680\,N^2
-5519/414720\,N^4$\\[1mm]
5 & $-1349303 /72900$ & $-\frac{75784701173}{1119744000}$ & $-181271/233280\,N^{-1} + 194389/405000\,N$\\
& & & $-262760809/116640000\,N^3$\\
& & & $- 61724321/1866240000\,N^5$\\
\hline
\end{tabular}
\end{center}
\caption{\label{tab:curvatureV}The coefficients $c_i$ of the leading logarithm $L^i_\phys$ up to $i=5$ for the
curvature $c_V$ for the physical cases $N=2$ and $N=3$ as well as for general $N$ in units of $M_\pi^2$.}
\end{table}

The formula in the massless case is much simpler.
The logarithm is now a bit more unique. We replace $\M^2$ by $-t$
and define
\begin{equation}\label{Kt}
 K_t \equiv \frac{t}{16\pi^2 F^2} \log\Big(-\frac{\mu^2}{t} \Big) \, .
\end{equation}
Taking the limit $M_\pi^2 \to 0$ (which implies $F_\pi \to F$), we get
from (\ref{resultFV}):
\begin{align}
\label{FVchiral}
F_V^0(t) = 1 &+ K_t (N/12) + K_t^2 (N^2/288) + K_t^3 (N/32 + 5/5184\,N^3)
\notag\\&
  + K_t^4 (-N^2/345600 + N^4/518400)
\notag\\&
  + K_t^5 (871/115200\,N + 545009/181440000\,N^3 + 126059/6531840000\,N^5)\,.
\end{align}

We close this section with giving the expansion for the radius and curvature
of the vector form factor defined by
\begin{equation}
 F_V(t) = 1 + \frac16 \langle r^2 \rangle_V t + c_V t^2 + \cdots\,.
\end{equation}
The coefficients $c_i$ for the expansion in physical quantities are given
in Tables~\ref{tab:radiusV} and~\ref{tab:curvatureV} in units of $M_\pi^2$.
The result up to two-loop order agrees with the LL extracted from the full
two-loop calculation~\cite{Bijnens:1998fm}. We do not present numerical results
for the vector form factor since these are dominated by large higher-order
contributions, see, e.g.,~\cite{Gasser:1983yg,Bijnens:1998fm}.

All the results presented in this section agree for $N=2$ up to fifth order
with the findings of~\cite{Bijnens:2012hf}.

\section{Scalar form factor}
\label{secsff}

\begin{table}
\small
\begin{center}
\begin{tabular}{|c|c|c|l|}
\hline
$i$ & $c_i$ for $N=2$ & $c_i$ for $N=3$ & $c_i$ for general $N$\\
\hline
1  & $-1$ & $-2/3$  &  $-2 N^{-1}$\\[1mm]
2  & $31/8$ & $569/72$  &   $23/2\,N^{-2} - 7/2 + 9/8\,N^2$\\[1mm]
3 & $65/6$  & $3205/81$  &  $-260/3\,N^{-3} + 133/3\,N^{-1} - 95/12\,N + 23/12\,N^3$\\[1mm]
4 & $76307/1152$ & $2330311/7776$ &  $19801/24\,N^{-4} - 5731/12\,N^{-2} + 1033/9$\\
& & & $-3331/288\,N^2 + 295/72\,N^4$\\[1mm]
5 & $375263/1152$ & $65426359/31104$ & $-189077/20\,N^{-5} + 82642/15\,N^{-3}
          - 1352663/1080\,N^{-1}$\\
& & & $+	 63967/360\,N - 35647/4320\,N^3 + 48487/5760\,N^5$\\
\hline
\end{tabular}
\end{center}
\caption{\label{tab:FS0}The coefficients $c_i$ of the leading logarithm $L^i_\phys$ up to $i=5$ for
the scalar form factor at zero momentum transfer $F_S(0)$ for the physical cases $N=2$ and $N=3$ as well as for general $N$.}
\end{table}

The (singlet) scalar form factor is defined by
\begin{align}
\langle \phi^a(p_f)| -j_0^s | \phi^a(p_i) \rangle =
 F_S\left[(p_f-p_i)^2\right]\,,
\end{align}
where again we have normalized the scalar current generator to one.
It is known fully in two- and three-flavour ChPT to 
one \cite{Gasser:1983yg,Gasser:1984ux} and
two loops \cite{Bijnens:1998fm,Bijnens:2003xg}. Here we calculate the
leading logarithms in the equal mass case to five loops.

As opposed to the vector case, the scalar form factor is not normalized to one,
such that we also need to specify the LL
expansion of $F_S(0)$. The coefficients $c_i$ for the expansion in terms of physical logarithms for $O_0 = 2B$ and
$O_\phys = F_S(0)$ are given in Table~\ref{tab:FS0}.
The momentum dependent part, $\tilde F_S(t)\equiv F_S(t)/F_S(0)$, can again
be expressed in terms of $\tilde t= t/M_\pi^2$ and the logarithm
defined in~(\ref{FFlog}). To fifth order we find
{\small
\begin{align}
\label{resultFS}
\tilde F_S&(t) = 1+ L_{\M} \Big[
           N/2\,\tilde t
          \Big]
       +  L_{\M}^2  \Big[
          \tilde t\,(-1/2 - 5/18\,N^2)
          +\tilde t^2\,43/144\,N^2
           \Big]\nonumber \\[1mm]&
       +  L_{\M}^3   \Big[
         \tilde t \,(5/6\,N^{-1} - 3/8\,N - 473/2592\,N^3)
         + \tilde t^2 (-91/216\,N - 227/1296\,N^3)
         + \tilde t^3\, 143/864\,N^3
          \Big]
\nonumber \\[1mm]
&       + L_{\M}^4   \Big[
          \tilde t \,(-13/6\,N^{-2} - 1751/1944 - 10529/5184\,N^2 - 2117/3456\,N^4)
\nonumber\\&
         + \tilde t^2 \,(4645/3888 + 245537/388800\,N^2 + 27103/259200\,N^4)
\nonumber\\&
         + \tilde t^3 \,(-196121/388800\,N^2 - 57061/388800\,N^4)
         +\tilde t^4\, (1129/57600\,N^2 + 580837/6220800\,N^4)
           \Big]
\nonumber \displaybreak[4]\\[1mm]&
         + L_{\M}^5   \Big[
          \tilde t \,(139/10\,N^{-3} + 327877/29160\,N^{-1} - 1550429/233280\,N
          - 28557851/4665600\,N^3
\nonumber\\&
          - 3800759/3110400\,N^5)
          + \tilde t^2 \,(-222149/58320\,N^{-1} - 858337/466560\,N
          \nonumber\\&
          + 410235883/233280000\,N^3 - 39345049/466560000\,N^5)
          \nonumber\\&
          + \tilde t^3 \,(324253/233280\,N - 7699463/38880000\,N^3 + 26029871/311040000\,N^5)
\nonumber\\&
           + \tilde t^4 \,(-1129/144000\,N - 357457/1296000\,N^3 - 18692191/186624000\,N^5)
\nonumber\\&
           + \tilde t^5 \,(315439/25920000\,N^3 + 48727189/933120000\,N^5)
          \Big]
\,.
\end{align}}

\begin{table}[t]
\small
\begin{center}
\begin{tabular}{|c|c|c|l|}
\hline
$i$ & $c_i$ for $N=2$ & $c_i$ for $N=3$ & $c_i$ for general $N$\\
\hline
1  & 6  & 9         &  $3 N$ \\[1mm]
2  & $-29/3$  & $-18$        &   $ -3 - 5/3\,N^2 $\\[1mm]
3  & $-581/54$  & $-1663/48$  &  $5 N^{-1} -9/4\,N - 473/432\,N^3$ \\[1mm]
4 & $-75301/648$ & $-2147363/5184$ &  $-13 N^{-2} - 1751/324 - 10529/864\,N^2 - 2117/576\,N^4$\\[1mm]
5  & $-\frac{5482247}{9720}$ & $-\frac{535098163}{186624}$ & $417/5\,N^{-3} + 327877/4860\,N^{-1} - 1550429/38880\,N$\\
& & & $- 28557851/777600\,N^3 - 3800759/518400\,N^5$\\
\hline
\end{tabular}
\end{center}
\caption{\label{tab:radiusS}The coefficients $c_i$ of the leading logarithm $L^i_\phys$ up to $i=5$ for the
the scalar radius $\langle r^2 \rangle_S$ for the physical cases $N=2$
and $N=3$ as well as for general $N$.}
\end{table}

\begin{table}[t]
\small
\begin{center}
\begin{tabular}{|c|c|c|l|}
\hline
$i$ & $c_i$ for $N=2$ & $c_i$ for $N=3$ & $c_i$ for general $N$\\
\hline
1  & 0 & 0        &  0\\[1mm]
2  & $43/36$ & $43/16$  &   $43/144\,N^2$\\[1mm]
3 & $-727/324$  & $-863/144$  &  $-91/216\,N - 227/1296\,N^3$\\[1mm]
4 & $4369/810$ & $2386939/155520$ &  $4645/3888 + 245537/388800\,N^2 + 27103/259200\,N^4$\\[1mm]
5 & $\frac{16871641}{2916000}$ & $\frac{1130937893}{55987200}$ & $-222149/58320\,N^{-1} - 858337/466560\,N$\\
& & & $+ 410235883/233280000\,N^3$\\
& & & $- 39345049/466560000\,N^5$\\
\hline
\end{tabular}
\end{center}
\caption{\label{tab:curvatureS}The coefficients $c_i$ of the leading logarithm $L^i_\phys$ up to $i=5$ for the
curvature $c_S$ for the physical cases $N=2$ and $N=3$ as well as for general $N$.}
\end{table}

As for the vector form factor, we can also give our result for the radius and the curvature, which are defined as
\begin{equation}
\tilde F_S(t) = 1 + \frac16 \langle r^2 \rangle_S t + c_S t^2 + \cdots\,.
\end{equation}
The coefficients $c_i$ for the expansion in physical quantities are given
in Tables~\ref{tab:radiusS} and~\ref{tab:curvatureS}.

All the results presented in this section agree for $N=2$ up to fourth order with the findings of~\cite{Bijnens:2010xg}.

\section{Meson-meson scattering}
\label{secpipi}

The amplitude for general meson-meson scattering is defined from
\begin{align}
\langle \phi^c(p_3) \phi^d(p_4) \text{out} | 
 \phi^a(p_1) \phi^b(p_2) \text{in} \rangle
	= i (2\pi)^4 \delta^4 (p_3 + p_4 - p_1 - p_2) M(s,t,u) \,,
\end{align}
with the Mandelstam variables
\begin{align}
s = (p_1+p_2)^2, \qquad t = (p_1-p_3)^2, \qquad u = (p_1-p_4)^2 \,.
\end{align}
It has been calculated at one-loop order in~\cite{Chivukula:1992gi}.
The two-loop calculation has been performed together with two other symmetry
breaking patterns in~\cite{Bijnens:2011fm}.

The structure of the $SU(N)$ meson-meson scattering amplitude has been
derived in full generality in \cite{Chivukula:1992gi,Bijnens:2011fm}.
It can be expressed in terms of two invariant amplitudes $B(s,t,u)$
and $C(s,t,u)$ as
\begin{align}
M(s,t,u) =\; &\left[\tr{T^a T^b T^c T^d}+\tr{T^a T^d T^c T^b}\right] B(s,t,u)
\nonumber\\
	&+\left[\tr{T^a T^c T^d T^b}+\tr{T^a T^b T^d T^c}\right] B(t,u,s)
\nonumber\\
	&+\left[\tr{T^a T^d T^b T^c}+\tr{T^a T^c T^b T^d}\right] B(u,s,t)
\nonumber\\
	&+\delta^{ab}\delta^{cd} C(s,t,u)+\delta^{ac}\delta^{bd} C(t,u,s)
					+\delta^{ad}\delta^{bc} C(u,s,t)\,.
\end{align}
The $T^a$ are the generators of $SU(N)$ normalized as $\tr{T^a T^b} = \delta^{ab}$.
Crossing symmetry implies
\begin{align}
	B(s,t,u) = B(u,t,s) \,, \qquad C(s,t,u) = C(s,u,t) \,.
	\label{pipicross}
\end{align}
In the case of $N=2$ the traces over four generators evaluate to products of Kronecker deltas such that the structure
of the amplitude is reduced to the well-known expression
\begin{align}
	M(s,t,u) = \delta^{ab}\delta^{cd} A(s,t,u)+\delta^{ac}\delta^{bd} A(t,u,s) +\delta^{ad}\delta^{bc} A(u,s,t) \,,
\end{align}
with
\begin{align}
	A(s,t,u) = C(s,t,u) + B(s,t,u) + B(t,u,s) - B(u,s,t)\,.
\end{align}
This is of course the structure of the $\pi \pi$ scattering amplitude in two-flavour ChPT.

We have calculated the LL contribution to the two invariant amplitudes to fifth order. In order to make the symmetries~(\ref{pipicross})
explicit, we have expressed $B(s,t,u)$ in terms of
\begin{align}
	\tilde t = t/M_\pi^2 \quad \text{and} \quad \tilde \Delta_{su} = (s-u)/M_\pi^2 \,,
\end{align}
and $C(s,t,u)$ in terms of
\begin{align}
	\tilde s = s/M_\pi^2 \quad \text{and} \quad \tilde \Delta_{ut} = (u-t)/M_\pi^2 \,.
\end{align}
We also express it in terms of the more general logarithm (\ref{FFlog}).
Our result for general $N$ reads
\newcommand{\phan}{\nonumber\\&\phantom{+L_\M}}
{\small
\begin{align}
	&\frac{F_\pi^2}{M_\pi^2} B(s,t,u)=1 - \tilde t/2
		+ L_\M [-2/3 (3 N^{-1} - N) - 5/12\,N \tilde t + 1/16\,N \tilde t^2 + 1/48\,N \tilde \Delta_{su}^2]\nonumber\\[1mm]
		&+ L_\M^2 [1/36 (684 N^{-2} + 12 + 29 N^2)  - 1/36 (54 N^{-2} + 156 + 17 N^2) \tilde t\phan
			 + 1/288 (486 + 29 N^2) \tilde t^2 - 1/288 (42 - 11 N^2) \tilde \Delta_{su}^2 - 5/1152 (54 + N^2) \tilde t^3\phan
		- 5/384 (6 + N^2) \tilde t \tilde \Delta_{su}^2
          ]\nonumber\\[1mm]
		&+ L_\M^3 [- 1/6480 (1188000 N^{-3} - 224160 N^{-1} - 54388 N - 7461 N^3)\phan
			+ 1/12960 (90720 N^{-3} + 40200 N^{-1} - 192278 N - 7391 N^3) \tilde t\phan
			- 1/51840 (33840 N^{-1} - 317952 N - 7127 N^3) \tilde t^2\phan
			+ 1/51840 (20400 N^{-1} - 17888 N + 3219 N^3) \tilde \Delta_{su}^2\phan
			- 1/51840 (56769 N + 502 N^3) \tilde t^3
			- 1/51840 (9403 N + 1646 N^3) \tilde t \tilde \Delta_{su}^2\phan
			+ 17/207360 (711 N + 8 N^3) \tilde t^4
			+ 1/207360 (4122 N + 653 N^3) \tilde t^2 \tilde \Delta_{su}^2\phan
			- 1/69120 (43 N - 29 N^3) \tilde \Delta_{su}^4]\nonumber\\[1mm]
      &+ L_\M^4 [- 1/777600 (- 1710720000 N^{-4} + 272984400 N^{-2} + 7733020 - 24562906 N^2\phan
			- 1393823 N^4)
			+ 1/3110400 (- 208396800 N^{-4} - 216817200 N^{-2} + 32764628\phan
			- 120616232 N^2 - 2240271 N^4) \tilde t
			+ 1/777600 (10376100 N^{-2} + 11270867\phan
			+ 12959626 N^2 + 141465 N^4) \tilde t^2
			+ 1/3110400 (- 9954000 N^{-2} + 4785940 - 1937608 N^2\phan
			+ 310421 N^4) \tilde \Delta_{su}^2
			- 1/12441600 (14175000 N^{-2} + 90102306 + 45489100 N^2\phan
			+ 198731 N^4) \tilde t^3
			- 1/12441600 (4725000 N^{-2} + 4860662 + 4068564 N^2\phan
			+ 725363 N^4) \tilde t \tilde \Delta_{su}^2
			+ 1/24883200 (33470280 + 9600665 N^2 + 52084 N^4) \tilde t^4\phan
			- 1/2764800 (281520 - 173014 N^2 - 27275 N^4) \tilde t^2 \tilde \Delta_{su}^2
			- 1/24883200 (528120\phan
			+ 125223 N^2 - 33181 N^4) \tilde \Delta_{su}^4
			- 1/24883200 (2694708 + 395685 N^2 + 1270 N^4) \tilde t^5\phan
			- 1/12441600 (418230 + 63783 N^2 + 7363 N^4) \tilde t^3 \tilde \Delta_{su}^2\phan
			- 1/8294400 (135360 + 25767 N^2 + 2642 N^4) \tilde t \tilde \Delta_{su}^4]\nonumber\\[1mm]
      &+ L_\M^5 [- 1/233280000 (7018928640000 N^{-5} - 894502656000 N^{-3} - 
         283189944960 N^{-1}\phan + 72507663308 N - 22041000184 N^3 - 690252879 N^5)
       +1/3265920000 (2508879744000 N^{-5}\phan + 2387314944000 N^{-3} - 
         1788779140400 N^{-1} + 55061393444 N - 302932752254 N^3 \phan- 3125842333 
         N^5) \tilde t
       +  1/1866240000 (54336096000 N^{-3} - 23570392560 N^{-1} + 
         4485329438 N \phan- 1556519119 N^3 + 304085556 N^5) \tilde \Delta_{su}^2 - 1/13063680000
         (1174116384000 N^{-3}\phan - 743306538800 N^{-1} - 1129493851946 N - 
         538364942807 N^3 - 3202580448 N^5) \tilde t^2 \phan
       + 1/26127360000 (40960080000 N^{-3} + 14810807200 N^{-1} - 
         46526628758 N \phan- 14697445505 N^3 - 2544291752 N^5) \tilde t \tilde \Delta_{su}^2 + 1/
         26127360000 (122880240000 N^{-3} \phan- 16992114720 N^{-1} - 1115282964722 N
          - 281556881291 N^3 - 615480176 N^5) \tilde t^3\phan
       + 1/14929920000 (529874640 N^{-1} - 971377232 N - 249290704 N^3
          + 44246931 N^5) \tilde \Delta_{su}^4 \phan+ 1/52254720000 (14905099440 N^{-1} -
         17891241888 N + 7135232732 N^3 \phan + 1114519563 N^5) \tilde t^2 \tilde \Delta_{su}^2 - 1/
         104509440000 (17859033360 N^{-1} - 964620068912 N\phan - 176725206936 N^3 -
         477061005 N^5) \tilde t^4
         - 1/52254720000 (3722683128 N + 675880271 N^3\phan + 64699403 
         N^5) \tilde t \tilde \Delta_{su}^4 - 1/26127360000 (4775058204 N + 535939563 N^3 +
         59878229 N^5) \tilde t^3 \tilde \Delta_{su}^2 \phan- 1/52254720000 (53926623792 N + 7406736715
          N^3 + 10009775 N^5) \tilde t^5\phan
        - 1/5971968000 (1608228 N + 172954 N^3 - 42207 N^5) \tilde \Delta_{su}^6
          + 1/41803776000 (35608896 N \phan+ 27012324 N^3 + 4741685 N^5) \tilde t^2 
         \tilde \Delta_{su}^4 + 1/13934592000 (86939676 N + 29291830 N^3 \phan+ 1761661 N^5) \tilde t^4
          \tilde \Delta_{su}^2 + 1/209018880000 (7967262600 N + 1179147888 N^3 + 2029543 N^5)
          \tilde t^6]
\end{align}}
and
{\small
\begin{align}
	&\frac{F_\pi^2}{M_\pi^2} C(s,t,u) = \nonumber\\
		&+ L_\M [2 N^{-2} + 3/8\, \tilde s^2 + 1/8\,\tilde \Delta_{ut}^2]\nonumber\\[1mm]
		&+ L_\M^2 [-2 (7 N^{-3} - N^{-1}) - 1/36 (108 N^{-1} - 37 N) \tilde s - 13/36\,N \tilde s^2\phan
			+ 5/24\,N \tilde \Delta_{ut}^2  + 55/192\,N \tilde s^3 + 5/192\,N \tilde s \tilde \Delta_{ut}^2
          ]\nonumber\\[1mm]
		&+ L_\M^3 [1/810 (112860 N^{-4} + 5580 N^{-2} - 1173 + 1022 N^2)\phan
			+ 1/3240 (66240 N^{-2} - 35174 + 1841 N^2) \tilde s
			+ 1/2592 (4104 N^{-2} + 9256 + 1661 N^2) \tilde s^2\phan
			+ 1/12960 (6840 N^{-2} + 3936 + 3811 N^2) \tilde \Delta_{ut}^2
			- 1/12960 (19044 + 3929 N^2) \tilde s^3\phan
			+ 1/6480 (1998 + 353 N^2) \tilde s \tilde \Delta_{ut}^2
			+ 1/23040 (3762 + 4177 N^2) \tilde s^4\phan
			+ 1/11520 (1254 + 49 N^2) \tilde s^2 \tilde \Delta_{ut}^2
			+ 1/23040 (418 + 13 N^2) \tilde \Delta_{ut}^4
		]\nonumber\\[1mm]
		&+ L_\M^4 [- 1/25920 (41990400 N^{-5} + 6609600 N^{-3} - 4545600 N^{-1} + 365524 N - 112385 N^3)\phan
			- 1/1555200 (342921600 N^{-3} - 85806200 N^{-1} + 26129734 N - 1244537 N^3) \tilde s\phan
			- 1/3110400 (21772800 N^{-3} - 31914800 N^{-1} - 11247384 N - 615995 N^3) \tilde s^2\phan
			- 1/622080 (1451520 N^{-3} + 90000 N^{-1} - 1035376 N - 245073 N^3) \tilde \Delta_{ut}^2\phan
			+ 1/6220800 (12042900 N^{-1} + 1797031 N + 2858510 N^3) \tilde s^3\phan
			- 1/2073600 (864300 N^{-1} - 2339087 N - 167184 N^3) \tilde s \tilde \Delta_{ut}^2\phan
			- 1/12441600 (10710972 N + 2610871 N^3) \tilde s^4
			+ 1/103680 (46775 N + 1366 N^3) \tilde s^2 \tilde \Delta_{ut}^2\phan
			+ 1/2488320 (197684 N + 4431 N^3) \tilde \Delta_{ut}^4
			+ 1/33177600 (3838644 N + 3531083 N^3) \tilde s^5\phan
			+ 1/19906560 (275724 N + 4553 N^3) \tilde s \tilde \Delta_{ut}^4
			+ 1/9953280 (285444 N + 5093 N^3) \tilde s^3 \tilde \Delta_{ut}^2
		]\nonumber\\[1mm]
		&+ L_\M^5 [
	1/81648000 ( 1754648179200 N^{-6} + 506658499200 N^{-4} - 280670106528 N^{-2}\phan
	     + 54081789480- 2534550216 N^2 + 999897691 N^4)
       + 1/1632960000 (4317509952000 N^{-4} \phan
       - 1246702968000 N^{-2} +33240243520 - 42063222570 N^2 + 13158613 N^4)  \tilde s\phan
       + 1/6531840000 (147311136000 N^{-4} + 93672190080 N^{-2}
       - 67579015880 + 41982479569 N^2 \phan + 3435837529 N^4) \tilde \Delta_{ut}^2 + 1/
         6531840000 (441933408000 N^{-4} - 524994234240 N^{-2} \phan + 414501298952 +
         85767984257 N^2 + 6718221609 N^4)  \tilde s^2
       + 1/4354560000 ( 21832020000 N^{-2} \phan -10349834500 +
         12661599885 N^2 + 429983342 N^4)  \tilde s \tilde \Delta_{ut}^2 \phan- 1/13063680000 (
         238082755680 N^{-2} + 224361657012 + 45828303299 N^2 \phan+ 836998070 N^4)
          \tilde s^3 
       + 1/26127360000 (2833329240 N^{-2} + 1543460180 + 5975744498 N^2
          \phan+ 97700803 N^4) \tilde \Delta_{ut}^4 + 1/26127360000 (16999975440 N^{-2}
          + 11020206360 + 32781452788 N^2\phan + 783206303 N^4)  \tilde s^2 \tilde \Delta_{ut}^2 + 1/13063680000 (
          12749981580 N^{-2} + 117034928874 \phan + 16745354989 N^2 + 4551556991 N^4)
          \tilde s^4 
       +  1/10450944000 (1480020120 + 1322536120 N^2\phan + 14756647 N^4) 
          \tilde s^3 \tilde \Delta_{ut}^2 + 1/20901888000 (1604443896 + 1525971760 N^2 + 18544639
         N^4)  \tilde s \tilde \Delta_{ut}^4 \phan- 1/34836480000 (37975069080 + 25358850464 N^2 +
         4923041771 N^4)  \tilde s^5\phan
       +  1/580608000 (20634624 + 4848765 N^2 + 66668 N^4)  \tilde s^4 \tilde \Delta_{ut}^2
          + 1/2322432000 (77354172 \phan+ 15213300 N^2 + 129899 N^4)  \tilde s^2 \tilde \Delta_{ut}^4 +
         1/52254720000 (118623420 + 27848862 N^2 \phan+ 216847 N^4) \tilde \Delta_{ut}^6 + 1/
         34836480000 (2583793620 + 2952657684 N^2 + 2122427329 N^4)  \tilde s^6
      ]\,.
\end{align}}
The five-loop contribution is of the same calculational complexity as the mass to the sixth order and as the latter, has only been checked in two of the four parametrizations in \eqref{params}. Up to fourth order and for $N=2$, the amplitude agrees with $A(s,t,u)$ from \cite{Bijnens:2010xg}.

It is well known that for $SU(2)$ $\pi\pi$ scattering, the amplitude can be decomposed into three amplitudes corresponding
to intermediate states of fixed isospin 0, 1, or 2. This decomposition can be generalized to arbitrary values of $N$,
where one finds seven different intermediate states. The corresponding amplitudes are obtained from the above invariant
amplitudes as~\cite{Bijnens:2011fm}
\begin{align}
\label{TISUN}
T_I =\; &2\left(N-\frac{1}{N}\right) [B(s,t,u) + B(t,u,s)]- \frac{2}{N}B(u,s,t)\nonumber\\
			&+ (N^2-1) C(s,t,u) + C(t,u,s) +  C(u,s,t)\; ,\nonumber\\[1mm]
T_S =\; &\left(N - \frac{4}{N}\right)[B(s,t,u) + B(t,u,s)]- \frac{4}{N} B(u,s,t)\nonumber\\
			&+ C(t,u,s) + C(u,s,t)\; ,\nonumber\\[1mm]
T_A =\; & N[-B(s,t,u)+B(t,u,s)]+C(t,u,s)-C(u,s,t)\; ,\nonumber\\[1mm]
T_\textit{SA} =\; & C(t,u,s)- C(u,s,t)\; ,\nonumber\\[1mm]
T_{AS} =\; & C(t,u,s) - C(u,s,t)\; ,\nonumber\\[1mm]
T_{SS} =\; & 2B(u,s,t) + C(t,u,s) + C(u,s,t)\; ,\nonumber\\[1mm]
T_{AA} =\; &-2B(u,s,t) + C(t,u,s) + C(u,s,t)\; .
\end{align}
The full scattering amplitude is then built up from these as
\begin{align}
	M(s,t,u)=\sum_J T_J(s,t,u) P_J\,,
\end{align}
where $P_J$ are the respective projection operators. Since their explicit form is rather lengthy, we do not reproduce it
here and refer the interested reader to~\cite{Bijnens:2011fm}. For $N=2$, the channels with $J = S, \textit{SA}, AS, AA$
do not exist and for $N=3$ the channel with $J=AA$ is not present. Each channel can be projected on partial waves by
\begin{align}
	T^J_\ell(s) = \frac{1}{64\pi} \int^1_{-1} d(\cos\theta) P_\ell(\cos\theta) T_J(s,t,u)\,.
\end{align}
\begin{table}[t]
\small
\begin{center}
\begin{tabular}{|c|c|c|c|c|c|}
\hline
$i$ & $a^I_0$ for $N=2$ & $a^I_0$ for $N=3$ & $a^S_0$ for $N=3$ & $a^{SS}_0$ for $N=2$ & $a^{SS}_0$ for $N=3$\\
\hline
1 & 9/2  & 358/51 & 59/21 & $-3/2$ & $-14/9$ \\[1mm]
2 & 857/42  & 28487/612 & 3505/252 & $-31/6$ & $-955/108$ \\[1mm]
3 & 153211/1512 & 7143269/22032 & 751735/9072 & $-7103/216$ & $-255265/3888$ \\[1mm]
4 & 41581/84 & 98674513/44064 & 26921179/51840 & $-7802/45$ & $-6097649/12960$ \\[1mm]
5$^*$ & $\frac{139816697}{56700}$	 & $\frac{165016031929}{10575360}$  & $\frac{228658804229}{65318400}$ &$-\frac{3326573}{3375}$&$-\frac{83190853}{23040}$ \\[1mm]
\hline
\end{tabular}
\end{center}
\caption{\label{tab:a0}The coefficients $c_i$ of the leading logarithm $L^i_\phys$ up to $i=5$ for the
$s$-wave scattering lengths for the physical cases $N=2$ and $N=3$.}
\end{table}
\begin{table}[t]
\small
\begin{center}
\begin{tabular}{|c|c|c|c|}
\hline
$i$ & $a^A_1$ for $N=2$ & $a^A_1$ for $N=3$ & $a^{SA}_1 = a^{AS}_1$ for $N=3$ \\
\hline
1 & 2  & 13/6 & $2/3$ \\[1mm]
2 & 791/36  & 941/36 & $50/9$  \\[1mm]
3 & 8528/81 & 665171/3888 & $25481/972$  \\[1mm]
4 & 2291903/3888 & 678064381/559872	 & $51822143/279936$ \\[1mm]
5$^*$ & 894986647/291600	 & 234732737339/27993600 &$3480221279/2799360$ \\[1mm]
\hline
\end{tabular}
\end{center}
\caption{\label{tab:a1}The coefficients $c_i$ of the leading logarithm $L^i_\phys$ up to $i=5$ for the
$p$-wave scattering lengths for the physical cases $N=2$ and $N=3$.}
\end{table}
Expanding these around threshold in powers of $q^2 = s/4 - M_\pi^2$ leads to the definition of the threshold parameters:
\begin{align}
	\text{Re}\ T^J_\ell(s) = q^{2\ell} ( a^J_\ell + q^2 b^J_\ell + q^4 c^J_\ell + \cdots ) \,,
\end{align}
where $a^J_\ell$ are the scattering lengths and $b^J_\ell$ the slope parameters. We have calculated all the
$s$- and $p$-wave scattering lengths. Note that for each channel, only one of the two partial waves is non-zero.
For $N=2$, only three channels contribute and the scattering length are more commonly denoted by
\begin{align}
	a_0^0 \equiv a^I_0 \,, \qquad a_1^1 \equiv a^A_1 \,, \qquad a^2_0 \equiv a^{SS}_0 \,.
\end{align}
For $N=3$, $T^{AA}$ is the only channel that vanishes, such that there are six scattering lengths. The tree-level
expressions for general $N$ as well as for the physical cases $N=2$ and $N=3$ are given by \cite{Chivukula:1992gi,Bijnens:2011fm}.
\begin{align}\renewcommand{\frac}[1]{\dfrac{#1}}
	a^{I,\tree}_0 &= \frac{M_\pi^2}{16 \pi F_\pi^2}
            \left( 2 N - \frac{1}{N} \right) &
		&\stackrel{N=2}{\Rightarrow} \frac{7 M_\pi^2}{32 \pi F_\pi^2}
& &\stackrel{N=3}{\Rightarrow} \frac{17 M_\pi^2}{48 \pi F_\pi^2} \,,
\nonumber\\
	a^{S,\tree}_0 &= \frac{M_\pi^2}{16 \pi F_\pi^2}
                 \left( N -\frac{2}{N} \right) &
                &&
		&\stackrel{N=3}{\Rightarrow} \frac{7 M_\pi^2}{48 \pi F_\pi^2} \,,
\nonumber\\
	a^{A,\tree}_1 &= \frac{M_\pi^2}{48 \pi F_\pi^2} N &
	&\stackrel{N=2}{\Rightarrow} \frac{M_\pi^2}{24 \pi F_\pi^2} &
        &\stackrel{N=3}{\Rightarrow} \frac{M_\pi^2}{16 \pi F_\pi^2} \,,
\nonumber\\
	a^{\textit{SA},\tree}_1 &= a^{AS,\tree}_1 = 0 &
        &&
        &\stackrel{N=3}{\Rightarrow} 0\,,
\nonumber\\
	a^{SS,\tree}_0 &= -\frac{M_\pi^2}{16 \pi F_\pi^2} &
	&\stackrel{N=2}{\Rightarrow} -\frac{M_\pi^2}{16 \pi F_\pi^2} &
        &\stackrel{N=3}{\Rightarrow} -\frac{M_\pi^2}{16 \pi F_\pi^2} \,,
\nonumber\\
	a^{AA,\tree}_0 &= \frac{M_\pi^2}{16 \pi F_\pi^2} \,.
\end{align}
The scattering lengths from channels that do not contribute for $N=2$ or $N=3$ have been omitted.
$a^{\textit{SA},\tree}_1$ and $a^{\textit{SA},\tree}_1$ vanish at tree level, but the higher-order contributions are
non-zero. 

\begin{figure}[t]
\begin{minipage}{0.49\textwidth}
\includegraphics[width=\textwidth]{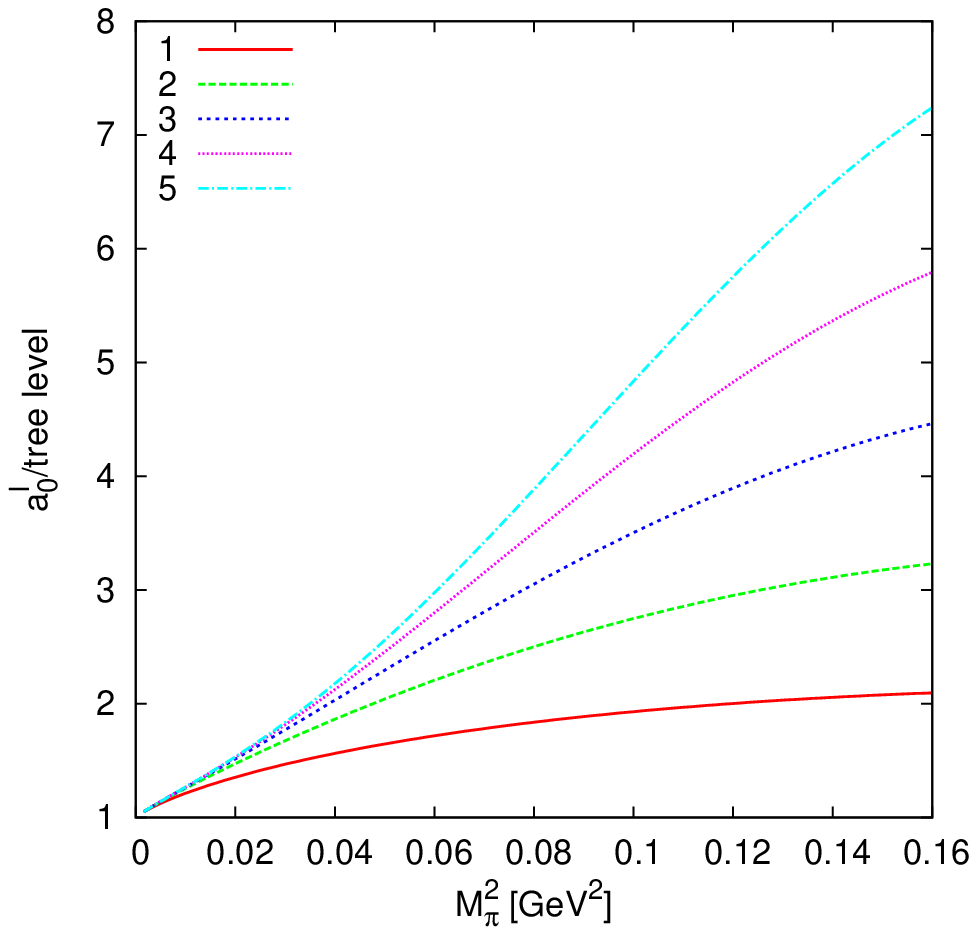}
\end{minipage}
\begin{minipage}{0.49\textwidth}
\includegraphics[width=\textwidth]{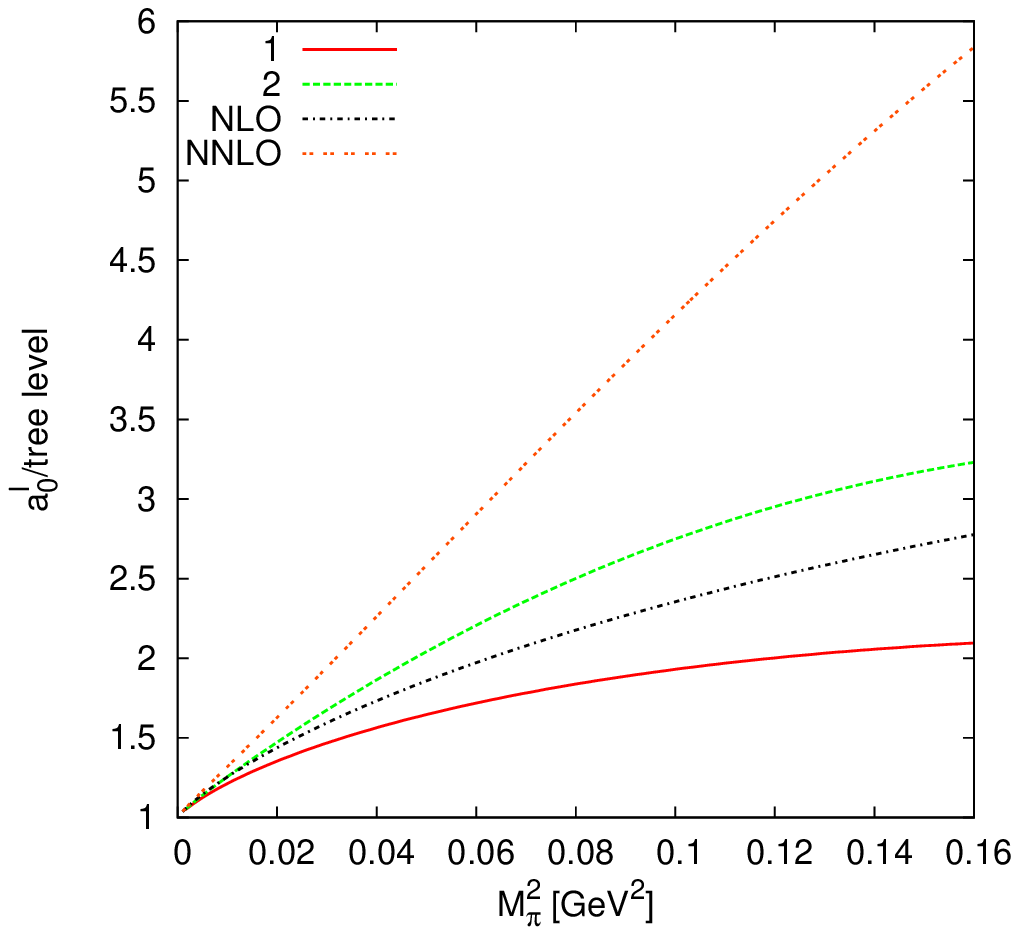}
\end{minipage}
\caption{\label{figpipi}
The contribution of the leading logarithms to $a^I_0/a^{I,tree}_0$
order by order for $F_\pi=0.0922$~GeV, $\mu =0.77$~GeV and $N=3$.
The left panel shows the expansion in $L_{\phys}$ keeping $F_\pi$ fixed.
The right panel shows the first two orders together with with the equivalent full
results of \cite{Bijnens:2011fm}, labeled NLO and NNLO.}
\end{figure}

As usual, the LL expansion can be written in the form
of~(\ref{defexpLphys}), where now $O_0 = a^{J,\tree}_\ell$.
There is, however, an exception: since the tree-level contribution to $a^\textit{SA}_1$ and $a^{AS}_1$ vanishes, the
series is written in these cases as
\begin{align}
	a^\textit{SA}_1 = a^{AS}_1 = \frac{M_\pi^2}{16 \pi F_\pi^2} ( c_1 L_\phys + c_2 L_\phys^2 + \cdots ) \,.
\end{align}
The corresponding coefficients for the two physical cases are listed in Table~\ref{tab:a0} for the $s$-wave and in
Table~\ref{tab:a1} for the $p$-wave scattering lengths.

Our results for the scattering lengths agree up to two-loops with~\cite{Bijnens:2011fm}. Furthermore, for $N=2$ the LL contributions to $a^I_0$ and $a^{SS}_0$ have been calculated up to fourth order
in~\cite{Bijnens:2010xg} and are in agreement with the present results.

We show the convergence of the leading logarithm part of the scattering length $a^I_0$ for $N=3$
in the left panel of Fig.~\ref{figpipi}. On the right side, we show the one- and two-loop
LL compared with the full calculation of \cite{Bijnens:2011fm}. One can see that the leading
logarithms are about half of the full correction.

\section{\texorpdfstring{$\gamma\gamma\to\pi\pi$}
{Gamma gamma to pi pi} and pion polarizabilities}
\label{ggpipi}

The process $\gamma\gamma\to\pi\pi$ has been calculated in ChPT
to one loop in~\cite{Bijnens:1987dc,Donoghue:1988eea}.
Already before the $p^6$ Lagrangian was explicitly known,
$\gamma\gamma\to\pi^0\pi^0$ has been worked out to two loops
in~\cite{Bellucci:1994eb} and $\gamma\gamma\to\pi^+\pi^-$
in~\cite{Burgi:1996mm,Burgi:1996qi}. Both calculations were redone
and the $\O(p^6)$ counter terms added explicitly
in~\cite{Gasser:2005ud,Gasser:2006qa}. While the leading contribution
to the charged process comes from tree-level diagrams, the neutral process
only starts at the one-loop level. In that case,
knowing at least the leading logarithms at higher orders is
a particularly welcome check for the convergence of the 
chiral expansion. Starting at the four-loop order, there is also
a doubly anomalous contribution to the amplitude.
However, it only affects sub-leading logarithms and is therefore not
relevant in the present context.

We have calculated the leading logarithms for the amplitude
with two different vectors to two different mesons in general but
the expressions are extremely lengthy. We thus restrict ourselves to the
simpler case where both vectors are coupling to the
current $\bar q \gamma_\mu T^c q$ and denote them as $\gamma^c$.
For the numerical results we treat only the case where the vectors correspond
to photons and are on-shell.

The $\gamma^c\gamma^c\to\pi^a\pi^b$ scattering amplitude is defined from the matrix
element
\begin{align}
\langle \pi^a(p_1) \pi^b(p_2) \text{out} | 
     \gamma^c(k_1) \gamma^c(k_2) \text{in} \rangle
= i (2\pi)^4 \delta^4 (p_1 + p_2 - k_1 - k_2) T^{abc}(s,t,u) \,,
\end{align}
with
\begin{align}
s = (p_1+p_2)^2\,, \qquad t = (p_1-k_1)^2\,, \qquad u = (p_1-k_2)^2 \,,
\end{align}
and
\begin{align}
	T^{abc}(s,t,u) = e^2 \epsilon_1^\mu \epsilon_2^\nu V_{\mu \nu}(s,t,u) \; .
\end{align}
The polarization vectors for the external vector are $\epsilon_1,\epsilon_2$
and we have added an overall coupling constant $e$ to the vectors.
We will consider the process for both vectors off-shell such that the amplitude
also depends on $k_1^2$ and $k_2^2$. Since both vectors carry the same flavour index,
the amplitude must satisfy
\begin{align}
k_1^\mu V_{\mu\nu} = k_2^\nu V_{\mu\nu}=0\,.
\end{align}
This follows from the Ward identity. If the vectors had different flavour there
could have been an equal time part in the Ward identity but it vanishes here.
As a result the amplitude can be
be decomposed into gauge invariant quantities as
\begin{align}
\label{defABCDE}
V^{\mu \nu}(s,t,u) = A(s,t,u) T_1^{\mu\nu} + B(s,t,u) T_2^{\mu\nu}
  + C(s,t,u) T_3^{\mu\nu} + D(s,t,u) T_4^{\mu\nu} + E(s,t,u) T_5^{\mu\nu}\,,
\end{align}
where
\begin{align}
	T_{1 \mu\nu} &= k_1 \cdot k_2 g_{\mu\nu} - k_{2 \mu} k_{1 \nu} \,,
\nonumber\\
	T_{2 \mu\nu} &= k_1 \cdot k_2 \Delta_\mu \Delta_\nu + \frac{1}{2} (t-u) k_{2 \mu} \Delta_\nu
				- \frac{1}{2} (t-u) \Delta_\mu k_{1 \nu} - \frac{1}{4} (t-u)^2 \; g_{\mu\nu} \,,
\nonumber \\
	T_{3 \mu\nu} &= k_1 \cdot k_2 k_{1 \mu} k_{2 \nu} - k_1^2 k_{2 \mu} k_{2 \nu}  
				- k_2^2 k_{1 \mu} k_{1 \nu} + k_1^2 k_2^2 g_{\mu\nu} \,,
\nonumber\\[1ex]	
	T_{4 \mu\nu} &= \begin{aligned}[t]
		&k_1 \cdot k_2k_{1 \mu} \Delta_\nu - k_1 \cdot k_2 \Delta_\mu k_{2 \nu} 
		- k_1^2 k_{2 \mu} \Delta_\nu + k_2^2 \Delta_\mu k_{1 \nu}\\
		&+ \frac{1}{2} (t-u) (k_1^2+k_2^2) g_{\mu\nu} 
		- \frac{1}{2} (t-u) k_{1 \mu} k_{1 \nu} - \frac{1}{2} (t-u) k_{2 \mu} k_{2 \nu} \,,
	\end{aligned}
\nonumber\\
	T_{5 \mu\nu} &= k_1^2 k_2^2 \Delta_\mu \Delta_\nu - \frac{1}{2} (t-u) k_1^2 \Delta_\mu k_{2 \nu}
        + \frac{1}{2} (t-u) k_2^2 k_{1 \mu} \Delta_\nu - \frac{1}{4} (t-u)^2 k_{1 \mu} k_{2 \nu} \,,
\end{align}
with $\Delta = p_1-p_2$. For $k_1^2 = k_2^2 = 0$, $T_{5 \mu\nu}$ is equivalent to $T_{3 \mu\nu}$, such that only four
independent quantities remain, which are identical to those given in \cite{Bellucci:1994eb} up to
normalization factors. For on-shell photons, one has in addition
$\epsilon_1 \cdot k_1 = \epsilon_2 \cdot k_2 = 0$, which only leaves $T_{1 \mu\nu}$ and $T_{2 \mu\nu}$.
The $T_{i \mu\nu}$ satisfy the identities
\begin{align}
	k_1^\mu T_{i \mu\nu} = k_2^\nu T_{i \mu\nu} = 0 \; ,
\end{align}
which can be readily checked using
\begin{align}
	- \Delta \cdot k_1 = \Delta \cdot k_2 = \frac{1}{2} (t-u) \; .
\end{align}

The flavour structure of the amplitude consists of all possible traces of
$T^a,T^b$ and $T^c$. But the total amplitude is symmetric under
$(p_1,a)\leftrightarrow(p_2,b)$ and
under $(k_1,\epsilon_1)\leftrightarrow(k_2,\epsilon_2)$.
Using charge conjugation\footnote{This is valid at least for the real part of the amplitude.} one can prove that the amplitude must be separately
invariant under $(a\leftrightarrow b)$.

\begin{figure}
\begin{center}\scalebox{0.85}{
\begin{picture}(160,160)(0,0)
    \SetWidth{0.7}
    \Photon(20,140)(63,97){5}{5}
    \Photon(20,20)(63,63){5}{5}
    \Line(140,140)(97,97)
    \Line(140,20)(97,63)
    \GCirc(80,80){25}{0.75}
    \Text(15,140)[cr]{$\gamma_1$}
    \Text(15,20)[cr]{$\gamma_2$}
    \Text(145,140)[cl]{$\phi^a$}
    \Text(145,20)[cl]{$\phi^b$}
  \end{picture}
\hspace{5mm}
	\begin{picture}(160,160)(0,0)
    \SetWidth{0.7}
    \Photon(20,140)(66,128){5}{4}
    \Photon(20,20)(66,32){5}{4}
    \Line(140,140)(94,128)
    \Line(140,20)(94,32)
    \Line(80,125)(80,35)
    \GCirc(80,125){15}{0.75}
    \GCirc(80,35){15}{0.75}
    \GCirc(80,80){15}{0.75}
    \Text(15,140)[cr]{$\gamma_1$}
    \Text(15,20)[cr]{$\gamma_2$}
    \Text(145,140)[cl]{$\phi^a$}
    \Text(145,20)[cl]{$\phi^b$}
  \end{picture}
\hspace{5mm}
	\begin{picture}(160,160)(0,0)
    \SetWidth{0.7}
    \Photon(20,140)(66,128){5}{4}
    \Photon(20,20)(66,32){5}{4}
    \Line(140,140)(95,35)
    \Line(140,20)(95,125)
    \Line(80,125)(80,35)
    \GCirc(80,125){15}{0.75}
    \GCirc(80,35){15}{0.75}
    \GCirc(80,80){15}{0.75}
    \Text(15,140)[cr]{$\gamma_1$}
    \Text(15,20)[cr]{$\gamma_2$}
    \Text(145,140)[cl]{$\phi^a$}
    \Text(145,20)[cl]{$\phi^b$}
  \end{picture}
}\end{center}
\caption{The three types of diagrams that contribute to
$\gamma \gamma \to \pi \pi$. The gray circles stand for the
sum over all one-particle-irreducible diagrams. Note that the
intermediate meson propagator in the two rightmost diagrams is off-shell.}
\label{fig:ggMMdiag}
\end{figure}
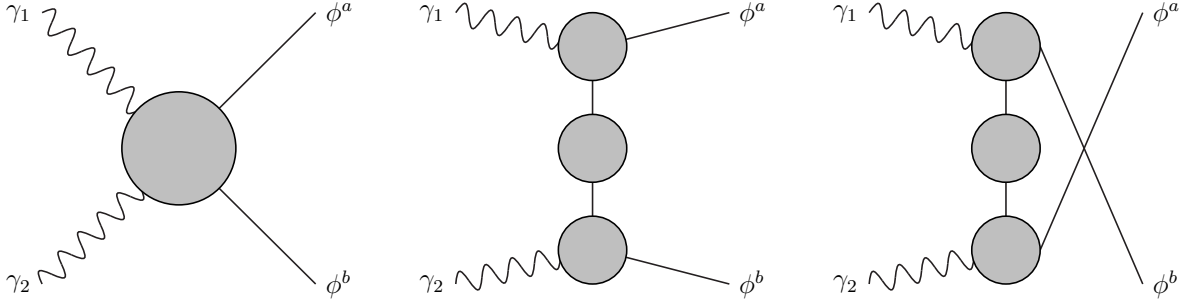
In contrast to all the earlier quantities discussed in this paper,
we must here consider several classes of diagrams in order to calculate
the physical
$\gamma \gamma \to \pi \pi$ amplitude. There are one-particle-reducible
contributions here other than those taken care of by wave function
renormalization.
The three needed types of diagrams are depicted in Figure~\ref{fig:ggMMdiag}. 
On the one hand, there is the direct contribution involving the
$\gamma\gamma\pi\pi$ leading-order vertex, on the other
hand, there are two types of diagrams involving the $\gamma\pi\pi$ vertex
twice. Loop contributions to the latter vertex have
been calculated already for the vector form factor and the results from there
can be reused. The corrections to the two-point function are incorporated by
using the physical propagator for off-shell momenta, which is given by
\begin{align}
	\frac{1}{1 + \overline \Sigma(p^2)} \frac{i}{p^2 - M_\pi^2} \quad \text{with} \quad
	\overline \Sigma(p^2) = \frac{\Sigma(p^2) - \Sigma(M_\pi^2)}{p^2 - M_\pi^2} \,,
\end{align}
where $\Sigma(p^2)$ denotes the self-energy of the meson. The contributions
from the diagrams involving a propagator can be written in a compact form
because these terms depend explicitly on the vector form factor. In particular,
because of the LSZ theorem, the residue of the amplitude when the intermediate
propagator is on-shell must contain the vector form factor twice.

We introduce a notation for the flavour traces that is a little
shorter to write:
\begin{align}
t_1 =& \langle T^a T^b T^cT^c\rangle+ \langle T^b T^a T^cT^c\rangle \,, &
t_2 =& \langle T^a T^c T^b T^c\rangle \,, &
\nonumber\\
t_3 =& \langle T^a T^b\rangle \langle T^c T^c\rangle \,, &
t_4 =& \langle T^a T^c\rangle \langle T^b T^c\rangle \,.&
\end{align}
That the traces in $t_1$ only appear together is due to the symmetry under $a \leftrightarrow b$.

A large part of the amplitude is contained in the
generalized Born amplitude because of the argument given above, with
\begin{align}
\label{BORN}
\epsilon_1^\mu \epsilon_2^\mu V_{\mu\nu}^{Born}
=& \left(t_1-2t_2\right) F_V(k_1^2)F_V(k_2^2)\Bigg[2\,\epsilon_1\cdot\epsilon_2
\nonumber\\
&+\frac{
-\Delta\cdot\epsilon_1 \,\Delta\cdot\epsilon_2
      +\Delta\cdot\epsilon_1 k_1\cdot\epsilon_2
      -k_2\cdot\epsilon_1\,\Delta\cdot\epsilon_2
      +k_2\cdot\epsilon_1\,k_1\cdot\epsilon_2}{t-M_\pi^2}
\nonumber\\
&+\frac{
 -\Delta\cdot\epsilon_1 \,\Delta\cdot\epsilon_2
      -\Delta\cdot\epsilon_1 k_1\cdot\epsilon_2
      +k_2\cdot\epsilon_1\,\Delta\cdot\epsilon_2
      +k_2\cdot\epsilon_1\,k_1\cdot\epsilon_2}{u-M_\pi^2}
\Bigg] \,.
\end{align}
The remainder of the amplitude now has no poles in the $t$ or $u$ channel.
The pole amplitude also has the commutator structure expected from tree-level
couplings to external vectors, visible in $T^{abc}$.
The expression for $F_V(t)$ in terms of the LLs can be found
in (\ref{resultFV}).
The generalized Born amplitude can be decomposed in the functions
defined in (\ref{defABCDE}), but is somewhat simpler in the form given
in (\ref{BORN}).

We write the full amplitude now as
\begin{align}
V_{\mu\nu} =& V_{\mu\nu}^{Born}
  +\overline A/M_\pi^2\, T_{1 \mu\nu}
  +\overline B/M_\pi^4\, T_{2 \mu\nu}
  +\overline C/M_\pi^4\, T_{3 \mu\nu}
	+\overline D/M_\pi^4\, T_{4 \mu\nu}
  +\overline E/M_\pi^6\, T_{5 \mu\nu} \,.
\end{align}
The factors of $M_\pi^2$ are introduced to make the functions dimensionless.
The partial amplitudes we write as functions of
\begin{align}
\tilde k_1 = k_1^2/M_\pi^2\,,\quad
\tilde k_2 = k_2^2/M_\pi^2\,,\quad
\tilde k_{12} = k_1\cdot k_2/M_\pi^2\,,\quad
\tilde \Delta_{tu} = (t-u)/M_\pi^2\,.
\end{align}
Each of the amplitudes we then write as
\begin{align}
\overline A = A^{(2)} L_\M^2+A^{(3)} L_\M^3 + A^{(4)} L_\M^4+\cdots\,.
\end{align}
The leading logarithms at one-loop order are already fully contained
in (\ref{BORN}). The two-loop leading logarithms are quite simple:
\begin{align}
B^{(2)} =\;&(1/72\, N^2+1/12) t_1
       -1/6\,t_2+1/9\,N t_3-1/36\,Nt_4\,,
\nonumber\\ 
A^{(2)} =\;& (\tilde k_{12} -2) B^{(2)},\quad
C^{(2)}= -B^{(2)},\quad D^{(2)}=E^{(2)} = 0\,.
\end{align}
The third-order expressions are still reasonable in full generality:
{\small
\begin{align}
A^{(3)} =\;& t_4   [
           1/216\, (30 \tilde k_{12}^2 + 9 \tilde k_2 \tilde k_{12} - 5 \tilde k_2^2 + 9 \tilde k_1 \tilde k_{12} - 2 \tilde k_1 
         \tilde k_2 - 5 \tilde k_1^2 + 64  \tilde k_{12} + 54  \tilde k_2 + 54  \tilde k_1 - 
         104 )
\nonumber\\&
          + 1/1296 (30 \tilde k_{12}^2 + 14 \tilde k_2 \tilde k_{12} - \tilde k_2^2 + 14 \tilde k_1 \tilde k_{12} + 4 \tilde k_1 
         \tilde k_2 - \tilde k_1^2 - 32  \tilde k_{12} + 50  \tilde k_2 + 50  \tilde k_1 + 88
         ) N^2
          ]
\nonumber\\&
       + t_3   [
         1/432 (30 \tilde k_{12}^2 + 9 \tilde k_2 \tilde k_{12} - 5 \tilde k_2^2 + 9 \tilde k_1 \tilde k_{12} - 2 \tilde k_1
         \tilde k_2 - 5 \tilde k_1^2 + 64  \tilde k_{12} + 54  \tilde k_2 + 54  \tilde k_1 -
         104)
\nonumber\\&
          + 1/2592 (30 \tilde k_{12}^2 - 31 \tilde k_2 \tilde k_{12} - 37 \tilde k_2^2 - 31 \tilde k_1 \tilde k_{12} - 50
         \tilde k_1 \tilde k_2 - 37 \tilde k_1^2 + 548  \tilde k_{12} - 1072) N^2
          ]
\nonumber\\&
       +  t_2   [
          1/1296 (210 \tilde k_{12}^2 + 21 \tilde k_2 \tilde k_{12} - 299 \tilde k_2^2 + 21 \tilde k_1 \tilde k_{12} + 58
          \tilde k_1 \tilde k_2 - 299 \tilde k_1^2 + 608  \tilde k_{12}
\nonumber\\&
    + 2042  \tilde k_2 + 2042 
          \tilde k_1 + 1832) N
          - 1/1296 (2 \tilde k_2 \tilde k_{12} + 8 \tilde k_2^2 + 2 \tilde k_1 \tilde k_{12} - \tilde k_1 \tilde k_2 + 8 
         \tilde k_1^2 - 26  \tilde k_{12}
\nonumber\\&
         - 52  \tilde k_2 - 52  \tilde k_1 + 12 ) N^3
          ]
\nonumber\\&
       +t_1   [
          - 1/1296 (60 \tilde k_{12}^2 - 3 \tilde k_2 \tilde k_{12} - 142 \tilde k_2^2 - 3 \tilde k_1 \tilde k_{12} + 32 
         \tilde k_1 \tilde k_2 - 142 \tilde k_1^2 
         + 208  \tilde k_{12} + 940  \tilde k_2
\nonumber\\&
         + 940 \tilde k_1 + 1072) N
          - 1/5184 (30 \tilde k_{12}^2 + 15 \tilde k_2 \tilde k_{12} - 13 \tilde k_2^2 + 15 \tilde k_1 \tilde k_{12} + 12
         \tilde k_1 \tilde k_2 - 13 \tilde k_1^2 - 40  \tilde k_{12}
\nonumber\\&
         + 160  \tilde k_2 + 160 \tilde k_1 + 184) N^3
          ]\,,
\nonumber\\
 B^{(3)} =\;& t_4   [
          1/216 (30 \tilde k_{12} + 29 \tilde k_2 + 29 \tilde k_1 - 20 )
          + 1/648 (15 \tilde k_{12} + 13 \tilde k_2 + 13 \tilde k_1 - 58 ) N^2
          ]
\nonumber\\&
 + t_3   [
           1/432 (30 \tilde k_{12} + 29 \tilde k_2 + 29 \tilde k_1 - 20 )
          + 1/2592 (30 \tilde k_{12} + 53 \tilde k_2 + 53 \tilde k_1 + 464 ) N^2
          ]
\nonumber\\&
  + t_2   [
          1/1296 (210 \tilde k_{12} + 293 \tilde k_2 + 293 \tilde k_1 - 1996 ) N
          + 1/432 (\tilde k_2 + \tilde k_1 - 10 ) N^3
          ]
\nonumber\\&
  + t_1   [
          - 1/1296 (60 \tilde k_{12} + 103 \tilde k_2 + 103 \tilde k_1 - 968 ) N
          - 1/5184 (30 \tilde k_{12} + 29 \tilde k_2 + 29 \tilde k_1 - 236 ) N^3
          ]
\nonumber \displaybreak[4]\\
 C^{(3)} =\;&
        t_4   [
          - 1/216 (34 \tilde k_{12} + 17 \tilde k_2 + 17 \tilde k_1 - 36 )
          - 1/324 (3 \tilde k_{12} + \tilde k_2 + \tilde k_1 - 27 ) N^2
          ]
\nonumber\\&
       + t_3   [
          - 1/432 (34 \tilde k_{12} + 17 \tilde k_2 + 17 \tilde k_1 - 44 )
          - 1/2592 (210 \tilde k_{12} + 121 \tilde k_2 + 121 \tilde k_1 + 452 ) N^2
          ]
\nonumber\\&
       +  t_2   [
          - 1/1296 (106 \tilde k_{12} + 77 \tilde k_2 + 77 \tilde k_1 - 1484 ) N
          - 1/1296 (\tilde k_2 + \tilde k_1 - 18 ) N^3
          ]
\nonumber\\&
       + t_1   [
           1/1296 (2 \tilde k_{12} + 13 \tilde k_2 + 13 \tilde k_1 - 676 ) N
          - 1/5184 (10 \tilde k_{12} + 7 \tilde k_2 + 7 \tilde k_1 + 200 ) N^3
          ]
\nonumber\\
 D^{(3)} =\;&
        t_4  \tilde\Delta_{tu} (
           5/216
          + 5/1296\, N^2
          )
       +  t_3   \tilde\Delta_{tu} (
           5/432
          + 5/2592\,N^2
          )
       + t_2   \tilde\Delta_{tu} (
          53/1296\, N
          + 1/2592 \, N^3
          )
\nonumber\\&
       + t_1  \tilde\Delta_{tu}  (
          - 19/1296 \, N
          - 1/864 \, N^3
          )
\nonumber\\
 E^{(3)} =\;&
        t_4   (
           7/54
          + 13/648\, N^2
          )
       + t_3   (
           7/108
          + 11/648\, N^2
          )
       + t_2 (
           23/162 \, N
          )
\nonumber\\&
       + t_1 (
          - 25/648\, N
          - 1/216\, N^3
          )\,.
\end{align}}
The fourth-order expression is very long. We therefore only quote the on-shell case
with $k_1^2=k_2^2=\epsilon_1\cdot k_1=\epsilon_2\cdot k_2=0$, where the amplitude is reduced to the contributions
from $\overline A$ and $\overline B$:
{\small
\begin{align}
A^{(4)}\! =\;&  t_4   [
           - 85/324 ( \tilde k_{12} - 2 ) N^{-1} + 1/777600 (129 \tilde k_{12} \tilde\Delta_{tu}^2 - 40908 \tilde k_{12}^3 + 12292  \tilde\Delta_{tu}^2 - 573518  \tilde k_{12}^2
\nonumber\\&
           + 3289918  \tilde k_{12} - 2678500 ) N - 1/1555200 (174 \tilde k_{12} \tilde\Delta_{tu}^2
          + 14452 \tilde k_{12}^3 + 1027  \tilde\Delta_{tu}^2
\nonumber\\&
         - 79808  \tilde k_{12}^2 - 67142  \tilde k_{12} - 75300 ) N^3
          ]
\nonumber\\&
       +  t_3   [
           43/324 ( \tilde k_{12} - 2 ) N^{-1}
          + 1/1555200 (5772 \tilde k_{12} \tilde\Delta_{tu}^2 + 346056 \tilde k_{12}^3 + 14281  \tilde\Delta_{tu}^2 -
         187274  \tilde k_{12}^2
\nonumber\\&
           + 1943624  \tilde k_{12} - 2697000 ) N
           + 1/3110400 (453 \tilde k_{12} \tilde\Delta_{tu}^2 + 37544 \tilde k_{12}^3 - 806  \tilde\Delta_{tu}^2
           + 6624\tilde k_{12}^2
\nonumber\\&
           + 1042176  \tilde k_{12} - 1985600 ) N^3
          ]
\nonumber\\&
       + t_2 [
         - 19/54 ( \tilde k_{12} - 2 ) N^{-2} - 1/388800 (1881 \tilde k_{12} \tilde\Delta_{tu}^2
         + 128988 \tilde k_{12}^3 + 663  \tilde\Delta_{tu}^2 + 128748  \tilde k_{12}^2
\nonumber\\&
           - 444098  \tilde k_{12}
  - 862700 )
          - 1/1555200 (4824 \tilde k_{12} \tilde\Delta_{tu}^2 + 326352 \tilde k_{12}^3 + 50627  \tilde\Delta_{tu}^2 - 
         1077658  \tilde k_{12}^2
\nonumber\\&
   - 5276492  \tilde k_{12} - 3115200 ) N^2
          - 1/62208 (39  \tilde\Delta_{tu}^2 + 280  \tilde k_{12}^2 - 5810  \tilde k_{12}
          + 6228 ) N^4
          ]
\nonumber\\&
       + t_1 [
          19/108 ( \tilde k_{12} - 2 ) N^{-2} + 1/777600 (1881 \tilde k_{12} \tilde\Delta_{tu}^2 + 128988 \tilde k_{12}^3 + 663  \tilde\Delta_{tu}^2 + 128748  \tilde k_{12}^2
\nonumber\\&
          - 460898  \tilde k_{12} - 829100 ) + 1/518400 (1139 \tilde k_{12} \tilde\Delta_{tu}^2 + 69072 \tilde k_{12}^3
          + 10597  \tilde\Delta_{tu}^2 - 253738  \tilde k_{12}^2
\nonumber\\&
           - 478512  \tilde k_{12} - 925400 ) N^2 + 1/6220800 (219 \tilde k_{12} \tilde\Delta_{tu}^2
          + 12512 \tilde k_{12}^3 + 5012  \tilde\Delta_{tu}^2 - 174348  \tilde k_{12}^2
\nonumber\\&
          + 48248  \tilde k_{12} - 20800 ) N^4
          ]
\nonumber\\
B^{(4)}\! =\;&
        t_4   [
        - 85/324\,  N^{-1} + 1/777600 (129 \tilde\Delta_{tu}^2 - 40908 \tilde k_{12}^2 + 564866  \tilde k_{12} - 362950 ) N
\nonumber\\&       
          - 1/777600 (87 \tilde\Delta_{tu}^2 + 7226 \tilde k_{12}^2 - 83402  \tilde k_{12} + 170475 
         ) N^3
          ]
\nonumber\\&
       +  t_3   [
         43/324 \, N^{-1} + 1/777600 (2886 \tilde\Delta_{tu}^2 + 173028 \tilde k_{12}^2 - 323231  \tilde k_{12} + 216900 ) N
\nonumber\\&
           + 1/3110400 (453 \tilde\Delta_{tu}^2 + 37544 \tilde k_{12}^2 - 65888  \tilde k_{12} + 812200 ) N^3
          ]
\nonumber\\&
       + t_2 [
       - 19/54 \, N^{-2}
          - 1/388800 (1881 \tilde\Delta_{tu}^2 + 128988 \tilde k_{12}^2 - 403776  \tilde k_{12} + 68650 
         )
\nonumber\\&
          - 1/777600 (2412 \tilde\Delta_{tu}^2 + 163176 \tilde k_{12}^2 - 1391027  \tilde k_{12} +
         4833750 ) N^2\!
          + 29/10368 (2  \tilde k_{12} - 29 ) N^4
          ]
\nonumber\\&
       + t_1   [
       19/108 \, N^{-2}
        + 1/777600 (1881 \tilde\Delta_{tu}^2 + 128988 \tilde k_{12}^2 - 403776  \tilde k_{12} + 51850 )
\nonumber\\& 
          + 1/1555200 (3417 \tilde\Delta_{tu}^2 + 207216 \tilde k_{12}^2 - 1310482  \tilde k_{12} + 4768800 ) N^2
\nonumber\\&
          +  1/6220800 (219 \tilde\Delta_{tu}^2 + 12512 \tilde k_{12}^2 - 162724  \tilde k_{12} + 691400 ) N^4
          ]
\end{align}}
The two-loop leading logarithms\footnote{Our calculation is with a charge
matrix with vanishing trace. However, the singlet part does not appear
in the lowest-order Lagrangian, hence we get the correct result for $N=2$
using $Q=\mathrm{diag}(1/2,-1/2)$ rather than $Q=\mathrm{diag}(2/3,-1/3)$.}
for $N=2$ agree with those
of~\cite{Gasser:2005ud,Gasser:2006qa} for $\overline A$ and $\overline B$.
Note that the $\bar l_i$ in the formulas given there also contain a logarithm denoted by $\ell$.

We can now use these results to find the polarizabilities. These are defined
from the helicity amplitudes,
\begin{align}
H_{++} =  \frac{\overline A}{M_\pi^2} +  \frac{4 M^2_\pi -s}{2M_\pi^4}\, \overline B\,,
\quad H_{+-} = \frac{2(M_\pi^4-tu)}{M_\pi^4 s}\, \overline B\,,
\end{align}
at fixed $t=M_\pi^2$:
\begin{align}
\frac{\alpha}{M_\pi}H_{+\mp}(s,t=M_\pi^2)
= (\alpha_1\pm\beta_1)+\frac{s}{12}(\alpha_2\pm\beta_2)+\O(s^2)\,.
\end{align}
The leading terms are the dipole, the next-to-leading terms the quadrupole polarizabilities.
They can then be expanded as
\begin{align} \label{polExp}
\alpha_i\pm\beta_i =\frac{\alpha}{16\pi^2 F_\pi^2 M_\pi}
\left(c_{i\pm}+\frac{M_\pi^2 d_{i\pm}}{16\pi^2 F_\pi^2}+\O(M_\pi^4)\right)\,.
\end{align}
At one-loop order the leading logarithms vanish. This was shown for $N=2,3$
in the earlier works \cite{Bijnens:1987dc,Donoghue:1988eea,Guerrero:1997rd}.
The two-loop LLs for $N=2$ can be extracted most easily from
the expressions in~\cite{Gasser:2005ud,Gasser:2006qa}.
The conclusion is that the only terms containing LLs are the $d_{1+}$
and from \cite{Gasser:2005ud,Gasser:2006qa} we get
\begin{align}
d_{1+}(\pi^0) = \frac{2}{9}\log^2(\mu^2/M_\pi^2)\,,\quad
d_{1+}(\pi^+)  = \frac{4}{9}\log^2(\mu^2/M_\pi^2)\,.
\end{align}
Alternatively we can write the expression for the polarizabilities using our
notation as
\begin{align}
\label{defcipol}
\alpha_1\pm\beta_1 =&\frac{\alpha}{M_\pi^3}\sum_{i} c_i L_\phys^i\,,
\nonumber\\
\alpha_2\pm\beta_2 =&\frac{\alpha}{M_\pi^5}\sum_{i} c_i L_\phys^i\,.
\end{align}
The resulting coefficients are given in Table~\ref{tabpol}. Note that they should not be
confused with the coefficients $c_{i\pm}$ in~\eqref{polExp}. We only quote the
results for $N=2$. The general $N$ results depend on how one
extends the charge matrix to more flavours.
\begin{table}[t]
\begin{center}
\begin{tabular}{|c|c|c|c|c|c|c|c|c|}
\hline
 &\multicolumn{4}{c|}{neutral}&\multicolumn{4}{c|}{charged}\\
\hline
$i$ &$\alpha_1+\beta_1$ & $\alpha_1-\beta_1$&$\alpha_2+\beta_2$ & $\alpha_2-\beta_2$
&$\alpha_1+\beta_1$ & $\alpha_1-\beta_1$&$\alpha_2+\beta_2$ & $\alpha_2-\beta_2$\\
\hline 
1 & 0      & 0        &0       &0      & 0      & 0      &0       &0      \\
2 & 2/9    & 0        &0       &0      & 4/9    & 0      &0       &0      \\
3 & 4/9    &$-8/9$    &20/9    &16/3   &337/81  &14/9    &$-10/9$ &$-352/27$ \\
4 & $\frac{203}{243}$ & $-\frac{155}{18}$&$\frac{4631}{675}$&$\frac{4694}{81}$&
$\frac{29239}{972}$&$\frac{1045}{54}$&$-\frac{114829}{2025}$&
$-\frac{4529}{27}$\\[1mm]
\hline
\end{tabular}
\end{center}
\caption{\label{tabpol}The coefficients $c_i$ of the leading logarithms
for the pion polarizabilities as defined in (\ref{defcipol}) for
the physical case $N=2$.}
\end{table}

In Table~\ref{tabpol2} we have given the numerical LL contributions together with
the full two-loop results in units of $10^{-4}~$fm$^3$ for the dipole
and $10^{-4}~$fm$^3$ for the quadrupole polarizabilities. 
\begin{table}[t]
\begin{center}
\begin{tabular}{|c|c|c|c|c|c|c|c|c|}
\hline
 &\multicolumn{4}{c|}{neutral}&\multicolumn{4}{c|}{charged}\\
\hline
$i$ &$\alpha_1+\beta_1$ & $\alpha_1-\beta_1$&$\alpha_2+\beta_2$ & $\alpha_2-\beta_2$
&$\alpha_1+\beta_1$ & $\alpha_1-\beta_1$&$\alpha_2+\beta_2$ & $\alpha_2-\beta_2$\\
\hline 
1 & 0     & 0      &0    &0   & 0    & 0   &0       &0      \\
2 & 0.11  & 0      &0    &0   & 0.23 & 0   &0       &0      \\
3 & 0.011 &$-0.022$&0.11 &0.26& 0.10 &0.039&$-0.056$&$-0.65$ \\
4 & 0.001 &$-0.011$&0.017&0.14& 0.037&0.024&$-0.14$ &$-0.42$\\
\hline
\cite{Gasser:2005ud,Gasser:2006qa} & 1.1 &$-1.9$ & 0.04 &37.6& 0.16 & 5.7 &$-0.001$& 16.2\\
\hline
\end{tabular}
\end{center}
\caption{\label{tabpol2}The numerical contribution of
the leading logarithms at each order to
the pion polarizabilities with the physical charged pion mass
$M_\pi = 139.57$~MeV, $F_\pi = 92.2$~MeV and $\mu=0.77$~GeV.
The units are $10^{-4}~$fm$^3$ for the dipole
and $10^{-4}~$fm$^5$ for the quadrupole polarizabilities.}
\end{table}
The results are somewhat mixed. For the phenomenologically
most relevant case, $\alpha_i-\beta_i$, the LL are always small and well below
the uncertainty quoted in \cite{Gasser:2005ud,Gasser:2006qa}. Also
$\alpha_1+\beta_1$ for the neutral pion is not much affected but
the other $\alpha_i+\beta_i$ obtain significant corrections to their actual
value. The estimate in \cite{Gasser:2005ud} of the $p^8$ contributions from
omega exchange to $\alpha_2+\beta_2$ neutral is $-0.25$ in the same units,
again much larger than the LL. The charged $\alpha_i+\beta_i$ obtain larger
relative corrections since the two-loop estimate of \cite{Gasser:2006qa} is very small.
Note that the chiral logarithm contributions in \cite{Gasser:2005ud,Gasser:2006qa}
also include nonleading logarithms which is why those numbers differ from
the ones in Table~\ref{tabpol2}.

\section{Conclusions}

In this work we extended the earlier work on leading logarithms in
effective theories and especially massive nonlinear sigma models to the
case of $SU(N)\times SU(N)/SU(N)$ with equal meson masses. 
We presented results for the leading
logarithms for up to 7 loops for the mass, decay constant, vacuum expectation value,
vector form factor, and scalar form factor, as well as for a number of 
quantities connected with meson-meson scattering and $\gamma\gamma\to\pi\pi$.
When applicable we have provided results for both 
physically relevant cases $N=2$ and $N=3$ and for general $N$.
In all cases we find reasonable convergence for $N=3$, while it is even better
for $N=2$.

All the results presented here have been checked through the use of several
parameterizations. Furthermore, we have compared to existing one- and two-loop
results. The higher orders could be checked for $N = 2$ by comparison with
our previous work on the $O(3)$ model.

We have not done a general study of how well the large $N$
limit works. But looking at the coefficients in the various tables, one notes
that the coefficients of the subleading terms are often larger than the
leading coefficients even though usually not by much. Some cases with
substantial corrections to the large $N$ limit can however be found.
In conclusion, the large $N$ limit is not typically a good approximation to
the full result but is significantly better than was found for the
$O(n)$ case, both due to the size of the coefficients and the fact that
the suppression is now in powers of $N^2$.

One of the motivations behind
this work was the hope that knowing many of the leading coefficients in $N$
would allow for an educated guess at the all-order series. We did not
succeed in this. Our work can serve as a starting point for future studies in this direction.

\section*{Acknowledgments}
S.L.\ is supported by a grant from the Swiss National Science Foundation
and K.K.\ by project MSM0021620859 of the Ministry of
Education of the Czech Republic and by a project UNCE No. 204020/2012 of
Charles University.
The Albert Einstein Center for Fundamental Physics is supported by the
``Innovations- und Kooperationsprojekt C-13'' of the ``Schweizerische
Universitätskonferenz SUK/CRUS''.
This work is supported in part by the European Community-Research
Infrastructure Integrating Activity ``Study of Strongly Interacting Matter''
(HadronPhysics3, Grant Agreement No. 283286)
and the Swedish Research Council grants 621-2011-5080 and 621-2010-3326.

\appendix

\section{Powers of \texorpdfstring{$N$}{N} in the results}
\label{appA}

The formulas and tables in the main text show a clear step of 2 in the
powers of the number of flavours $N$ that show up. In this appendix we show that
this must be the case to all orders. The same is true for the colour factors in
QCD with $N_c$-colours due to gluonic contributions but
we are not aware of a simple published proof even though it might be well
known in the QCD perturbation theory community.

If we had a theory with the symmetry breaking pattern $U(N)\times U(N)/U(N)$
we would have $N^2$ Goldstone bosons and the proof could be done using
't Hooft's double line notation with the only difference that the lines indicate flavour
of the quarks rather than colour. The argument in \cite{'tHooft:1973jz}
in the purely gluonic theory with surfaces with handles goes through in the same way
and one concludes that only positive powers of $N$ should appear and that adding
a handle changes the power by two.

Since in our case we do not have $N^2$ particles but rather $N^2-1$, we cannot simply
take over the arguments from \cite{'tHooft:1973jz}. Those still determine the
highest power of $N$ at each loop order, though.

In the main text we determined the leading logarithms from one-loop diagrams
only. But the leading logarithms are in principle also determined from the
diagrams with largest number of loops at any order. We use the fact
that in these diagrams, all vertices come from the single trace lowest-order
Lagrangian. The proof immediately generalizes if external fields are
included. We give the proof for one-particle-irreducible diagrams only
but the generalization should be fairly clear.

For a given one-particle-irreducible diagram, the number of loops $N_L$, vertices $N_V$, and propagators $N_P$ satisfy
the relation
\begin{eqnarray}
\label{loopsvertices}
N_L &=& N_P-N_V+1 \,.
\end{eqnarray}
For the parametrization $U_1$ in (\ref{params}), exactly one flavour trace appears for each vertex, such that the total number of
traces in the diagram is $N_V$. For each of the $N_P$ propagators there is a sum over a flavour index, which can
be evaluated using the relations
\begin{equation}
\label{traces}
\langle T^a A \rangle\langle T^a B\rangle =
\langle AB\rangle -\frac{1}{N}\langle A\rangle\langle B \rangle
,\qquad
\langle T^a A  T^a B\rangle =
\langle A\rangle\langle B\rangle -\frac{1}{N}\langle A B \rangle\,.
\end{equation}
We denote the total number of traces in a term by $N_{tr}$ and the power of $N$ by $N_N$.
The sum of the two is abbreviated by $N_{trN} = N_{tr} + N_N$. Each time the first one of the
relations~(\ref{traces}) is used, $N_{trN}$ is reduced by one. The first term of the second relation,
on the other hand, adds one to $N_{trN}$, while the second term again subtracts one.
Positive powers of $N$ can be generated by $\langle \unitmatrix\rangle = N$.

After all propagator flavour traces have been removed, a diagram with $N_V$ vertices can contain terms with 
$N_{trN} = N_V-N_P, N_V-N_P + 2, \ldots$. The minimal value is achieved if $N_{trN}$ has
been decreased by one $N_P$ times. If the first term of the second relation in~(\ref{traces}) has entered exactly once,
one gets instead $N_{trN} = N_V - (N_P-1) + 1$ and so on. From~(\ref{loopsvertices}) then follows that $N_{trN}$ is
odd (even) for even (odd) $N_L$. The tree-level one-particle-irreducible diagrams clearly satisfy this since they
contain no powers of $N$ and one flavour trace.

For a given number of loops, the lowest number that can occur is $N_{trN} = 1-N_L$ using~(\ref{loopsvertices}).
The highest requires a little more work because not all contractions can increase $N_{trN}$.
In a one-particle-irreducible loop diagram with $N_V$ vertices, we need to use the first relation in~(\ref{traces}),
which only lowers $N_{trN}$, at least $N_V-1$ times since at least that many operations
have the $T^a$ in different traces. So the maximum is 
\begin{equation}
N_{trN}=N_V-(N_V-1)+(N_P-(N_V-1)) = N_P-N_V+2= 1+N_L\,.
\end{equation}
This coincides with the maximum power derived with the double line method
of \cite{'tHooft:1973jz}.

\bibliographystyle{JHEPmod}
\bibliography{SUNLogs}

\end{document}